\documentclass[pra,aps,showpacs,floatfix,twocolumn,nofootinbib]{revtex4-1}
\usepackage{amsmath}
\usepackage{amsfonts}
\usepackage{amssymb}
\usepackage{revsymb}
\usepackage{graphicx}
\usepackage{bm}
\usepackage{dcolumn}
\usepackage{multirow}
\usepackage{pdfpages}

\newcommand{\fref}[1]{Fig.~\ref{#1}}
\newcommand{\sref}[1]{Sec. \ref{#1}}
\newcommand{\eref}[1]{Eq.~(\ref{#1})}
\newcommand{\tref}[1]{Table~\ref{#1}}

\def\ket#1{|\,#1 \,\rangle}

\def\vect#1{\mathbf{#1}}

\makeatletter
\AtBeginDocument{\let\LS@rot\@undefined}
\makeatother

\begin{document}
\title{Dynamic polarizability measurements in $^{176}$Lu$^+$}
\author{K. J. Arnold$^{1}$}
\email{cqtkja@nus.edu.sg}
\author{ R. Kaewuam$^{1}$}
\author{T. R. Tan$^{1,2}$}
\author{S. G. Porsev$^{3,4}$}
\author{M. S. Safronova$^{3,5}$}
\author{M. D. Barrett$^{1,2}$}
\email{phybmd@nus.edu.sg}
 \affiliation{
 $^1$Centre for Quantum Technologies, 3 Science Drive 2, 117543 Singapore \\
 $^2$Department of Physics, National University of Singapore, 2 Science Drive 3, 117551 Singapore \\
 $^3$Department of Physics and Astronomy, University of Delaware, Newark, Delaware 19716, USA \\
 $^4$Petersburg Nuclear Physics Institute of NRC ``Kurchatov Institute'', Gatchina, Leningrad District 188300, Russia \\
 $^5$Joint Quantum Institute, National Institute of Standards and Technology and the University of Maryland, College Park, Maryland 20742, USA
 }
\begin{abstract}
We measure the differential polarizability of the $^{176}$Lu$^+$ $^1\!S_0\leftrightarrow {^3}\!D_1$ clock transition at multiple wavelengths.  This experimentally characterizes the differential dynamic polarizability for frequencies up to 372\,THz and allows an experimental determination of the dynamic correction to the blackbody radiation shift for the clock transition.  In addition, measurements at the near resonant wavelengths of 598 and 646\,nm determine the two dominant contributions to the differential dynamic polarizability below 372\,THz.  These additional measurements are carried out by two independent methods to verify the validity of our methodology. We also carry out a theoretical calculation of the polarizabilities 
using the hybrid method that combines the configuration interaction (CI) and the coupled-cluster approaches, incorporating for the first time quadratic non-linear terms and partial triple excitations in the coupled-cluster calculations. The experimental measurements of the $|\langle ^3D_1|| r || ^3P_J\rangle|$ matrix elements provide high-precision benchmarks for this theoretical approach.
\end{abstract}
\maketitle

The differential scalar polarizability, $\Delta\alpha_0(\omega)$, of a clock transition is an important quantity to determine.  The dc value $\Delta\alpha_0(0)$ quantifies the blackbody radiation (BBR) shift, and contributes to micromotion shift assessments in ion-based clocks.  The variation of the polarizability over the BBR spectrum determines the so-called dynamic correction to the BBR shift~\cite{safronova2010black}, and the value at the clock frequency quantifies sensitivity to probe-induced ac Stark shifts.

For the  $^{176}$Lu$^+$ $^1S_0\leftrightarrow {^3}D_1$ transition at 848\,nm, the recent measurement of $\Delta\alpha_0(\omega)$ at $10.6\,\mu\mathrm{m}$ inferred an exceptionally small BBR shift of $-1.36(9)\times 10^{-18}$ at 300\,K~\cite{kja2018static}.  As the measurement was carried out at a frequency that is fairly central to the BBR spectrum, the assessment is insensitive to the true dc value of $\Delta\alpha_0(\omega)$ and its variation over the BBR spectrum.  Nevertheless it is still of interest to make an experimental assessment as $\Delta\alpha_0(0)$ can factor into planned assessments of the dc polarizability of the $^1S_0\leftrightarrow {^3}D_2$  and $^1S_0\leftrightarrow {^1}D_2$ clock transitions at 804 and 577\,nm, respectively.

The accuracy of the BBR assessment for the 848-nm transition relies on the small measured value of $\Delta\alpha_0(\omega)$ at 10.6
$\,\mu$m; a modest fractional error in a small number is still a small number.  This is not the case for the other two clock transitions in $^{176}$Lu$^+$.  For these two transitions micromotion-induced shifts can be used to determine $\Delta\alpha_0(0)$ as done in \cite{dube2014high}. For $^{176}$Lu$^+$ this can be elegantly done by measuring frequency ratios within the same apparatus.  In this case many systematics are common mode and the difference in the ratio with and without micromotion depends only on the micromotion amplitude, which can be accurately characterized from micromotion sidebands, and the difference in $\Delta\alpha_0(0)$ for the two transitions.  Assessment of $\Delta\alpha_0(0)$ for the 804- and 577-nm transitions by comparison to the 848-nm clock would then be limited by the small contribution from the 848-nm transition.

In light of the above considerations, $\Delta\alpha_0(\omega)$ for the 848-nm transition has been measured at six optical frequencies corresponding to the approximate wavelengths 1560, 987, 848, 804, 646, and 598\,nm.  All measurements, together with previous measurements at $10.6\mu\mathrm{m}$ \cite{kja2018static}, are then used to formulate a model for $\Delta\alpha_0(\omega)$ over the measurement window providing an estimate of $\Delta\alpha_0(0)$ and a reassessment of the BBR shift.

Measurements at 598 and 646\,nm determine the dominant pole contributions to $\Delta\alpha_0(\omega)$ at lower frequencies and largely determines the frequency dependence below 372\,THz ($\lambda=804\,\mathrm{nm}$).  The 598 and 646 measurements are independently verified using an alternative technique based on the comparison of ac Stark shifts and scattering rates \cite{ca_scattering2015}.  With this technique, the dependence on laser intensity factors out and provides a consistency check for the more conventional approach that involves characterizing the beam intensity~\cite{rosenband2006blackbody,YbPeik2016}.

The paper is organized into three main sections.  The first section details the experimental and theoretical methodologies, the measurements made, and compares theoretical and experimental results for the matrix elements and polarizabilities.  The second section develops a suitable model for $\Delta\alpha_0(\omega)$ based on a theoretical understanding of the atomic structure and supported by the measurements.  An independent assessment based on the single pole approximation \cite{YbPeik2016} is used for comparison as a means to check for modeling dependencies.  The final section applies the results to the BBR assessment.
\section{Polarizability measurements}
The experimental methodology employed is similar to that reported in previous work~\cite{kja2018static}. Linearly polarized light is focused on the ion to induce an ac Stark shift. This shift is measured on either the optical transition, $\ket{^1S_0,F$=$7,m_F$=$0} \leftrightarrow \ket{^3D_1,7,0}$, or the microwave transition $\ket{^3D_1,7,0} \leftrightarrow \ket{^3D_1,6,0}$. The optical transition is realized as an average of the $\ket{^1S_0,7,\pm1} \leftrightarrow \ket{^3D_1,7,0}$ transitions probed by Rabi spectroscopy with typical $\pi$-times of 5-20 ms. The ac Stark shift is measured by an interleaved servo technique~\cite{tamm2009stray,kja2018static}.  The laser intensity is determined from {\it in situ} 2D profiling of the beam at the ion and power measurements using a calibrated detector.

\subsection{Experimental setup and optical power characterization}
Experiments are performed in the same linear Paul trap used for previous work~\cite{dean2018}.  The trap consists of two axial endcaps separated by 2 mm and four rods arranged on a square with sides 1.2$\,$mm in length. All electrodes are made from 0.45$\,$mm electropolished copper-beryllium rods. Radial confinement is provided by a 16.8$\,$MHz radio-frequency (rf) potential applied to a pair of diagonally opposing electrodes via a helical quarter-wave resonator. A dc voltage applied to the other pair of diagonally opposing electrodes ensures a splitting of the transverse trapping frequencies. The endcaps are held at 8 V to provide axial confinement. The measured trap frequencies of a single Lu$^+$ are $(\omega_x, \omega_x, \omega_z) \approx 2\pi \times (608, 560, 134)\,\mathrm{kHz} $, with the trap axis along $z$.

\begin{figure}
\includegraphics[width=\columnwidth]{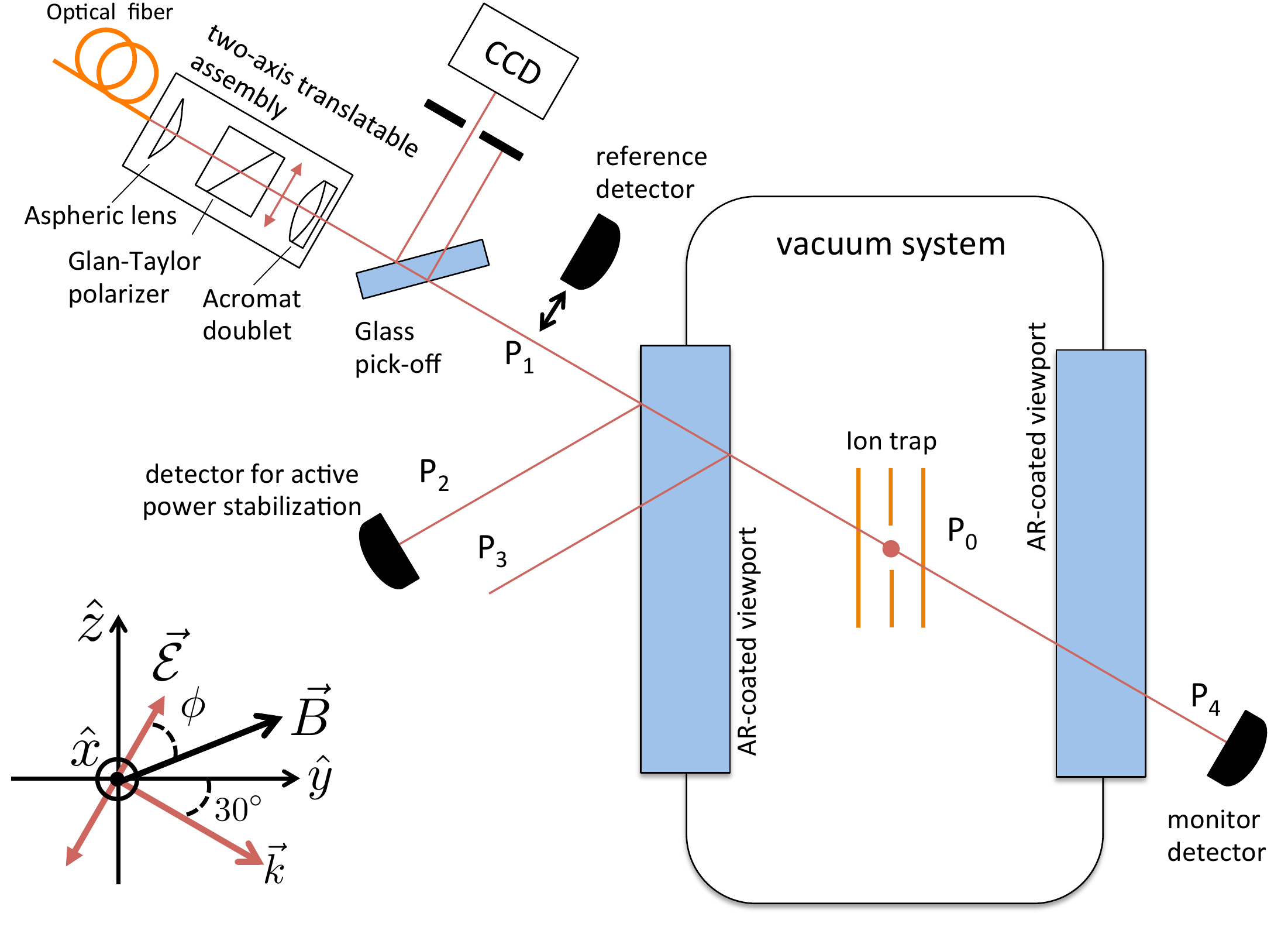}

\caption{Schematic of experimental setup. The light shifting laser is delivered to the experiment by optical fiber. The optics assembly to collimate and then focus the light onto the ion is able to be displaced along both axes orthogonal to the beam direction using motorized translation stages. A CCD camera monitors the beam displacement. Power measurement at the points $P_1$ to $P_4$ are used to infer the power at ion, $P_0$, as described in the main text. The laser direction $\vec{k}$ is approximately $30^\circ$ from normal incidence with respect to the viewport. The externally applied magnetic field $\vec{B}$ is rotated in the $yz$-plane to form an angle $\phi$ with respect to the linear laser polarization $\vec{\mathcal{E}}$.}
\label{fig:schematic}
\end{figure}

The optical setup for the ac-Stark shift laser is shown schematically in \fref{fig:schematic}. The light is delivered to the experiment on a single mode optical fiber. An assembly consisting of an aspheric lens to collimate the fiber output, a Glan-Taylor to set the polarization, and an achromat doublet to focus onto the ion, is mounted on a motorized two-axis translation stage. The exact optical components of this assembly are changed as needed to be suitable for the laser wavelength used (1560, 987, 848, 804, 646, or 598\,nm).  The reflection from the first surface of the fixed-position glass pick-off is captured by a charge-coupled device (CCD) to characterize the movement of the stage during beam profiling.

The first reflection from the vacuum viewport, $P_2$ in \fref{fig:schematic}, is used to actively stabilize the optical power by feedback onto an acousto-optic modulator (AOM) before the optical fiber. When the active stabilization is engaged, the reading on the monitor power meter at $P_4$ is repeatable to the lowest significant display digit over a day, and measurements of the ac-Stark shift indicate fractional power instability less than $10^{-3}$ (see Supplemental Material).  To determine the optical power at the ion, $P_0$, the vacuum viewport transmission, $T=(P_1-P_2-P_3)/P_1$,  and the ratio of reference and monitor detectors readings,  $r = P_\mathrm{ref}/P_\mathrm{mon}$,  are measured while the active stabilization is disengaged.  The power at the ion with the stabilization engaged is then $P_0 =  r T P_4$, with an uncertainty determined by the calibration accuracy of the reference detector and the statistical uncertainty in $r$ and $T$.  At every laser wavelength used, the reference detector has been calibrated at a nearby wavelength by the National Metrology Centre (NMC) in Singapore with certified 2$\sigma$ uncertainty of 1.5\%. Further details and measurement data related to power uncertainty assessment are given in the Supplemental Material. 

\subsection{Beam profiling and intensity characterization}
In order to determine the laser intensity, the beam is profiled by measuring the position-dependent ac Stark shift $\delta f(x^\prime,y^\prime)$ induced on the $\ket{^1S_0,7,0} \leftrightarrow \ket{^3D_1,7,0}$ clock transition as the beam is displaced by motorized translation stages (\fref{fig:schematic}) in a two-dimensional space ($x^\prime$,$y^\prime$) orthogonal to the laser direction.   We define a normalization constant
\begin{equation}
\label{eq:normconst}
C = \frac{\delta f_{\mathrm{max}}}{\iint \delta f(x^\prime,y^\prime) dx^\prime dy^\prime}
\end{equation}
where $\delta f_{\mathrm{max}}$ is the peak ac Stark shift and $C$ has units of m$^{-2}$.   The peak beam intensity is then $I_0 = C P_0$. A useful length scale to parameterize the mode is the effective waist $w_\mathrm{e} = \sqrt{2/(\pi C)}$, which corresponds to the waist of a Gaussian beam with the same normalization constant $C$.

\begin{figure*}
\includegraphics[width=\textwidth]{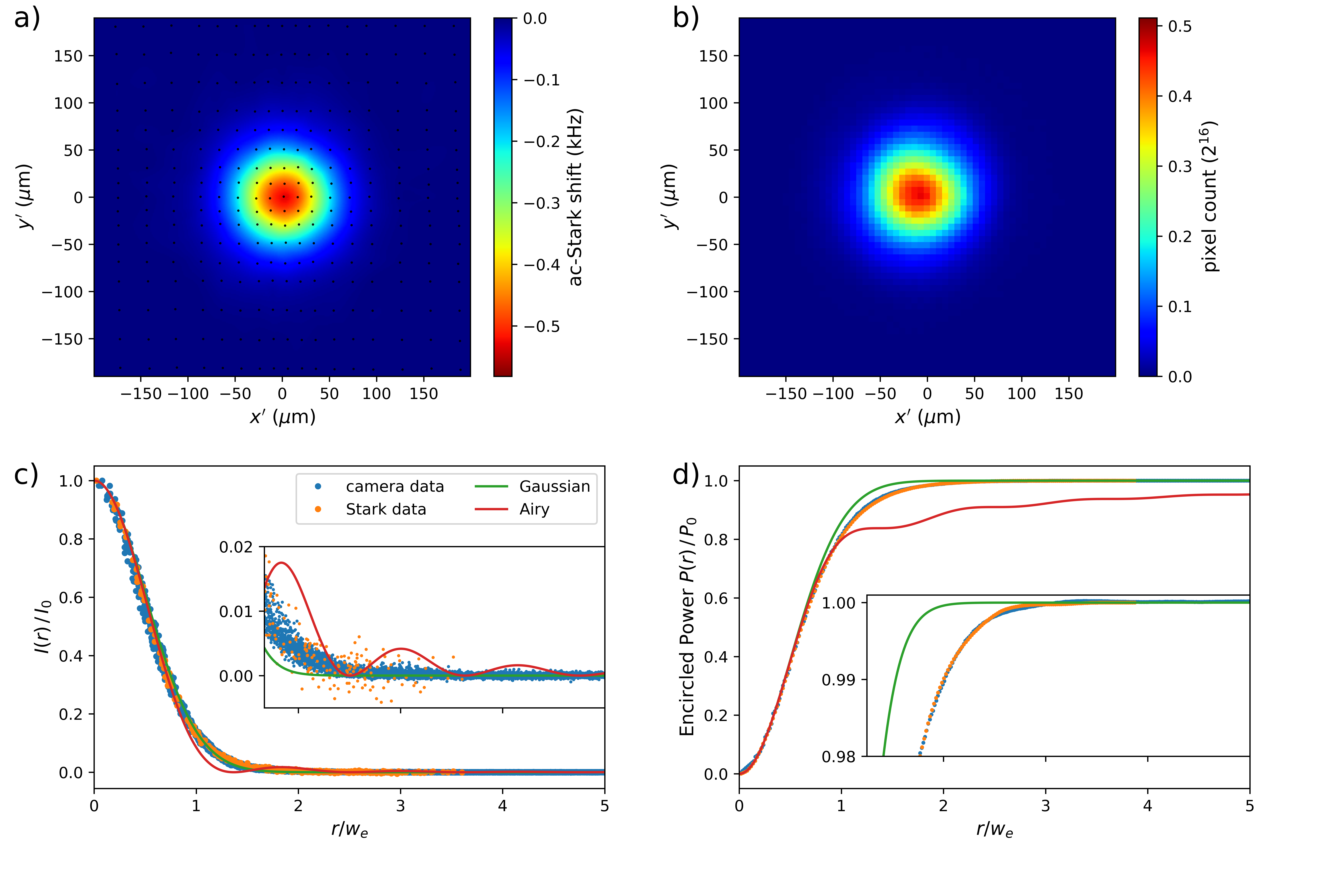}
\caption{Beam profiling data for the 598 nm laser. (a) Cubic-spline interpolation to ac-stark shift measurements. Black points indicate the measurement positions as determined by the CCD camera. (b) Beam profile captured on a low-noise camera outside the chamber at the focal plane of the ion. (c) Measured intensity as a function of radial distance from the beam center for the Stark shift (orange) and camera (blue) data. (d) Fraction of encircled power within radius $r$. (c-d) Gaussian (green) and Airy (red) test functions with the same normalization $C$ as the measured profile for comparison. Inset plots show magnified views for clarity.}
\label{fig:profile}
\end{figure*}

After each movement of the translation stages, the beam center position is determined by a 2D Gaussian fit to an image captured by the fixed position CCD camera shown in \fref{fig:profile}. The beam position determined by the CCD camera has $\pm$150$\,\mathrm{nm}$ repeatability, but is observed to drift by approximately $\sim3\,\mu$m over the course of one day, correlated with the ambient lab temperature. For a typical profile scan over a 300 $\mu$m square grid, the rms positioning error of the stages is $\sim1\,\mu$m as assessed by the CCD camera.  The beam displacement as measured by the camera is used for evaluating the beam profiles. Supporting data for the positioning accuracy is given in the Supplemental Material.

The ac-Stark shift profile for the 598 nm laser is shown in \fref{fig:profile}a as a representative dataset.   Here Rabi spectroscopy with a 9$\,$ms $\pi$-time is used. The measurement at each position is the average value after 20 interleaved servo updates, where one update occurs after 160 interrogations alternately with and without the Stark shift beam present. Before starting each measurement, the servo is run for 5 iterations to lock onto the Stark-shifted line to avoid servo error in the averaged value. For the data in \fref{fig:profile}a, the peak Stark shift is -528.8 Hz and the projection noise limited uncertainty at each position is 1.0 Hz.  The mode function is approximated by a cubic spline interpolation, and integration over the square data region yields a normalization constant of $C = 117.1 (3)\,\mathrm{mm}^{-2}$, corresponding to $w_{e} = 73.73(9)\,\mu$m.  The statistical uncertainty of $C$ is determined by a bootstrapping method where new data is generated by a Monte Carlo method allowing for variation due to (i) the projection noise in each measurement, (ii) an overall position offset of the coordinates with respect to the measure profile, and (iii) beam pointing drift over the duration of the profile measurement.  Three profiles were taken for both 646 and 804 nm and the repeatability was consistent with the estimated uncertainty as discussed in the Supplemental Material.   Only a single profile was taken for each of the other wavelengths. Datasets for other wavelengths are provided in the Supplemental Material.

A potential source of systematic error with this methodology is beam power not captured within the data region.  To illustrate, consider two test functions in \fref{fig:profile}(c-d) with the same normalization as our measured profile: a Gaussian (green line) and Airy distribution (red line) with respective intensity distributions
$$\frac{I(r)}{I_0} = e^{\frac{-2r^2}{w_\mathrm{e}^2}} \quad\mathrm{and} \quad \frac{1}{2} \left(\frac{w_\mathrm{e}}{r} J_1\left(\frac{r \sqrt{8}}{w_\mathrm{e}}\right)\right)^2,$$
 where $J_1$ is the Bessel function of the first kind of order one. The Airy function, a realistic optical profile resulting from uniform illumination of a circular aperture, has a significant fraction of power distributed over regions of large $r/w_e$. Even a partial contribution of the Airy distribution to the laser profile, which could result from either beam clipping or focusing aberration, for example, would be undetectable within the projection noise at the tails of the measured profile, yet still result in significant error for a data region extending to $r/w_e = 3.5$ (\fref{fig:profile}d). Considering the measured ac Stark shifts (orange points, \fref{fig:profile}c), even the modest deviation of the observed profile from a Gaussian at the tails of the distribution (inset \fref{fig:profile}c) requires the data region be extended from $r/w_e = 1.5$ to 2 in order to achieve 99\% power capture (inset \fref{fig:profile}d).

As an additional check for low level beam intensity distributed over a larger area, the beam was independently profiled with a low readout noise camera outside of vacuum. With the same optical assembly used for the experiment (\fref{fig:schematic}), including a glass pick-off and an identical vacuum viewport, beam profile images were captured at several positions around the focal plane. The camera image which had normalization nearest ($C = 118.8\,\mathrm{mm}^{-2}$) to the measured Stark profile is shown in \fref{fig:profile}b. From the camera data (blue points) in \fref{fig:profile}c, we see good agreement with the Stark data within the profiled region ($r/w_e < 3.5$) and no significant intensity beyond $r/w_e > 3.5$. Based on the camera data, the $360\,\mu$m square grid used for the Stark profile captures $\gtrsim$99.8\% of the power.

The normalization constants determined from the profiles at all wavelengths are summarized in \tref{tab:norms} with supporting data in the Supplemental Material. For the case of 1560 nm, the CCD camera is not sensitive to this wavelength and therefore stage movements could not be monitored. Additional uncertainty due to stage positioning was included in the bootstrapping method to assess the uncertainty contribution.  With the exception of the 646 and 598 nm laser profiles, the estimated errors due to power outside the profiling region are $1-2\%$. The $C$ values given in \tref{tab:norms} are corrected for this effect with the full size of the correction added to the uncertainty budget. A detailed evaluation of the uncertainties is given in the Supplemental Material.

\begin{table}
\caption{Results of beam profiling for all laser wavelengths $\lambda$. $l$ is the half-width of the square grid used for the profile scaled to the effective beam waist $w_e$. $P_\mathrm{loss}$ is the fraction of power estimated to be outside the profile data region. $C$ is the normalization constant as defined in \eref{eq:normconst}. Uncertainties are given in parentheses.}
\label{tab:norms}
\begin{ruledtabular}
\begin{tabular}{llcrlrl}
\multicolumn{1}{l}{$\lambda$ (nm)}
&\multicolumn{1}{l}{$l/w_e$} & \multicolumn{1}{c}{$P_\mathrm{loss}$}
& \multicolumn{2}{c}{$C$ (mm$^{-2}$)}& \multicolumn{2}{c}{$w_e$ ($\mu$m)} \\
\hline \\ [-0.3pc]
598&2.5&0.2\%&116.8&(1.1)&73.83&(35)\\
646&3.0&0.1\%&399.5&(2.7)&39.92&(13)\\
804&1.7&1.3\%&293.9&(2.6)&46.43&(21)\\
848&1.6&1.6\%&268.3&(5.3)&48.7&(5)\\
987&1.7&1.5\%&271.4&(4.8)&48.4&(4)\\
1560&1.8&1.1\%&84.2&(1.2)&87.0&(6)\\
\end{tabular}
\end{ruledtabular}
\end{table}

It is noted that even though the spatial mode is filtered by an optical fiber and focused with optics that have minimal spherical aberration, fitting to gaussian models is found to be insufficient for determining the peak intensity at the 1\% level. For example, if an elliptical gaussian distribution including TEM$_{0,0}$, TEM$_{0,1}$, and TEM$_{1,0}$ modes is used, as in \cite{YbPeik2016}, we find the normalization $C$ is consistently overestimated by 3-5\% for both the Stark shift and camera profile data at all wavelengths compared to the methodology employed here.

\subsection{Accurate assessment of the 646 and 598 poles}
Given the potential to mischaracterize the beam intensity, it would be advantageous to have an independent measurement to validate the methodology.  This can be done by conducting measurements near-to-resonant with a contributing transition.  For Lu$^+$, ideal candidates are the $^3D_1 \leftrightarrow  {^3}P_0$ and  $^3D_1 \leftrightarrow {^3}P_1$ transitions at 646 and 598\,nm, respectively.  Sufficiently near to the pole, the polarizability is, to a good approximation, determined by the single pole.  Additionally, there can be a measurable scattering rate, which is proportional to the Stark shift and the linewidth of the transition.  The ratio of Stark shift to scattering rate is then independent of the laser intensity. As demonstrated in \cite{ca_scattering2015}, this can provide an accurate assessment of the corresponding matrix element and hence polarizability.

\subsubsection{646 pole via polarizability}
The intensity of the Stark shift-inducing laser at 646 nm is actively stabilized with a peak intensity at the ion of $I_0 = 1.942 (21)\,\mathrm{W}\mathrm{cm}^{-2}$, as assessed by the methods of the previous sections. The laser is linearly polarized with the magnetic field aligned to the beam propagation axis ($\phi=90^\circ$) and detuned by $\Delta_0/2\pi = -241.7(2)\, \mathrm{GHz}$ from the $\ket{^3D_1,7,0}\leftrightarrow \ket{^3P_0,7,0}$ transition. All other transitions combined are estimated to contribute less than $<0.2\%$ to the differential dynamic polarizability of the clock transition at this detuning. To a good approximation, the Stark shift, $\hbar \delta_{0}$, of the $\ket{^3D_1,7,0}$ state is
\begin{equation}
\label{eq:646starkshift}
\delta_{0} = \frac{1}{6} \frac{\Omega_0^2}{4 \Delta_0}
\end{equation}
where $\Omega_0 = \frac{e a_0}{\hbar}\sqrt{\frac{2 I_0}{\epsilon_0 c}} \langle ^3D_1|| r || ^3P_0\rangle$. The measured shift of $\delta_{0}/2\pi = -846.5(3)\,\mathrm{Hz}$ at the position of peak intensity yields the matrix element:
\begin{equation}
|\langle ^3D_1|| r || ^3P_0\rangle| = 1.432 (8)\,\, {\rm a.u.}
\end{equation}

\subsubsection{646 pole via the scattering rate to stark shift ratio}
\begin{figure}
\includegraphics[width=\columnwidth]{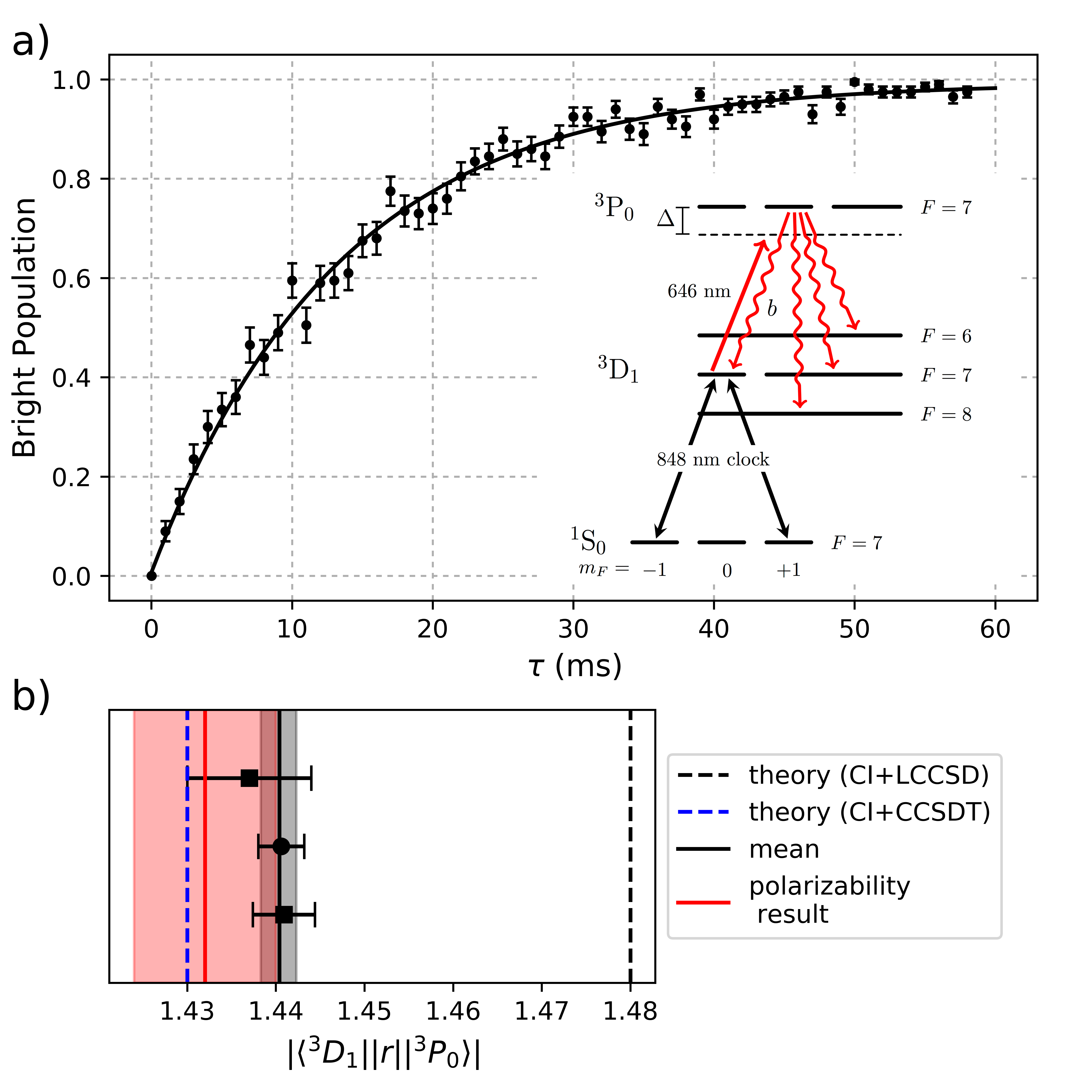}
\caption{a(inset) Schematic of energy levels. Population initially prepared in $\ket{^3D_1,7,0}$ is pumped out by a detuned 646 nm laser with circular polarization. The 848 nm clock laser is used for shelving to measure the population of $\ket{^3D_1,7,0}$ and to measure the ac Stark shift induced by the 646 nm laser. (a) Population pumped out of $\ket{^3D_1,7,0}$ after a 646 nm pulse of duration $\tau$. Black line is a fit to \eref{eq:646model}. (b) Comparison of the matrix element determined from the polarizability method (red line) and the scattering rate to Stark shift ratio (black line) from multiple experiments (black circles/squares). Shaded regions represent respective uncertainties. The black dashed line is the theory value from previous work~\cite{luprop2016} and the blue dashed line is a new theory value from the method applied in this work, \sref{sec:theory}. }
\label{fig:646scattering}
\end{figure}

The 646 laser for this measurement is derived from the detection and cooling laser but frequency offset to a detuning of $\Delta_0/2\pi \sim - 1\, \mathrm{GHz}$ from the $\ket{^3D_1,7,0} \leftrightarrow \ket{^3P_0,7,0}$ transition.   The optical power is actively stabilized but the absolute intensity at the ion is not accurately determined.  The beam propagates in the direction of the magnetic field ($\phi=90^\circ$) and has circular polarization ($\sigma^+$ coupling). This polarization ensures that once the atom Raman scatters out of $\ket{^3D_1,7,0}$ it cannot return to this state (\fref{fig:646scattering}a inset).  An atom prepared in the $\ket{^3D_1,7,0}$ state scatters via the $\ket{^3P_0,7,+1}$ state at the rate
\begin{equation}
\label{eq:646rate}
R_0 =  \frac{\Gamma_0}{6} \frac{\Omega_0^2}{4 \Delta_0^2}
\end{equation}
where $\Gamma_0$ is radiative decay rate of the $^3P_0$ state. From the ratio of \eref{eq:646starkshift} and \eref{eq:646rate}, one finds
\begin{equation}
\label{eq:646gamma}
\Gamma_0 = \frac{R_0 \Delta_0}{\delta_{0}}
\end{equation}
 where $R_0$, $\Delta_0$, and $\delta_{0}$ are all readily measurable quantities without characterization of the laser intensity.

The experimental procedure to measure $R_0$ is:
\begin{enumerate}
\item Repeat optical pumping into $\ket{^3D_1,7,0}$ ($\sim95\%$), shelving to $\ket{^1S_0,7,-1}$ on the clock transition, and detection of $^3D_1$ population until the atom is detected dark. This prepares the atom in  $\ket{^1S_0,7,-1}$ with $\sim99.8\%$ fidelity.
\item Shelve $\ket{^1S_0,7,-1}$ back to $\ket{^3D_1,7,0}$ with probability $P_s \approx 0.99$.
\item Apply detuned 646 laser for duration $\tau$
\item Shelve remaining $\ket{^3D_1,7,0}$ population to $\ket{^1S_0,7,+1}$ with probably $P_s$
\item Detect $^3D_1$ population
\end{enumerate}

The measured bright population after a pulse length of $\tau$ is then:
\begin{equation}
\label{eq:646model}
p(\tau) = P_s (1-P_s e^{-R_0 (1-b) \tau}),
\end{equation}
where $b$ is the fraction of Rayleigh scattering events back to $\ket{^3D_1,7,0}$ and is equal to $\frac{1}{6}$ for the states considered (\fref{fig:646scattering}a inset).

\fref{fig:646scattering}a shows the result of a typical preliminary experiment run which is fit to \eref{eq:646model} with $P_s$ and $R_0$ as free parameters. We acquire statistics on $P_s$, $R_0$, and $\delta_{0}$ from three interleaved experiments: (i) measure the bright population after preparation but without a 646 pulse to determine $P_s$ (ii) measure the population after a 646 pulse of fixed duration to determine $R_0$ from \eref{eq:646model}, and (iii) measure the ac-Stark on the $\ket{^3D_1,7,0}$ state using the 848-nm clock transition. A representative data set and statistical analysis corresponding to one of the points is given in the Supplemental Material.

The decay rate $\Gamma_0$ is determined from \eref{eq:646gamma} which is related to the matrix element by:
\begin{equation}
\Gamma_0 = \frac{\omega_0^3 e^2 a_0^2}{3 \pi \epsilon_0 \hbar c^3} |\langle ^3D_1|| r || ^3P_0\rangle|^2,
\end{equation}
where $\omega_0$ is the resonant transition frequency.

 \fref{fig:646scattering}b shows the results of measurements from three consecutive days with combined measurement time of approximately 35 hours. Black squares were taken at a detuning $\Delta_0 = 2\pi\times -989.46(10)\, \mathrm{MHz}$ and the black circle at $-1119.46(10)\, \mathrm{MHz}$.

The weighted mean result of the three experiments is
\begin{equation}
|\langle ^3D_1|| r || ^3P_0\rangle| = 1.440 (2)\,\, {\rm a.u.},
\end{equation}
indicated by the black line in \fref{fig:646scattering}b. For comparison, the red line is the result from the polarizability measurement, and the dashed lines are theoretical matrix elements from different methodologies discussed in \sref{sec:theory}. The results from the two experimental methodologies agree to within one standard error of the largest uncertainty.

Our initial results from the two methodologies were in significant disagreement. The source of the discrepancy was found to be the contribution of the amplified spontaneous emission (ASE) when using diode laser sources near to a resonance. For the polarizability measurement at $\Delta_0/2\pi = -241.7(2)\, \mathrm{GHz}$, we used a diffraction grating to filter the ASE before the optical fiber going to the experiment. This increased the measured Stark shift by 2.9(6)\% compared to no filtering and for the same laser intensity at ion. For the scattering rate measurement, a Fabry-P\`erot resonator was used to suppress ASE and undesired spurious spectral components near resonance (see Supplemental Material).

\subsubsection{598 pole via polarizability measurement}
The intensity of the Stark shift-inducing laser at 598 nm is actively stabilized with a peak intensity at the ion of $I_0 = 34.4 (6)\,\mathrm{mW}\mathrm{cm}^{-2}$. The laser frequency is referenced to an optical frequency comb and set to a detuning of $\Delta_1/2\pi = -1097.0(1)\,\mathrm{MHz}$ with respect to the $\ket{^3D_1,7,0} \leftrightarrow \ket{^3P_1,6,0}$ transition. The polarization is linear and aligned parallel to the externally applied magnetic field ($\phi = 0$). The Stark shift, $\hbar \delta_1$, on the $\ket{^3D_1,7,0}$ state is given by
\begin{equation}
\label{eq:598stark}
\delta_{1} = \frac{\Omega_1^2}{4}\left( \frac{4}{45} \frac{1}{\Delta_1} + \frac{7}{90} \frac{1}{\Delta_1-\omega_{68}}\right),
\end{equation}
where $\omega_{68} = 2\pi\times 52.832\,2\,\mathrm{GHz}$ is the separation of the $^3P_1$ $F$=$6$ and $F$=$8$ hyperfine levels, and $\Omega_1 = \frac{e a_0}{\hbar}\sqrt{\frac{2 I_0}{\epsilon_0 c}} \langle ^3D_1|| r || ^3P_1\rangle$. The second term in \eref{eq:598stark} is due to coupling from the  $\ket{^3D_1,7,0} \leftrightarrow \ket{^3P_1,8,0}$ transition and contributes 1.7\% to the total Stark shift. Since the laser is $\pi$ polarized there is no contribution from the $^3P_1$ $F$=$7$ level.
At the position of peak intensity, we measure a shift of $\delta_1/2\pi=-1318(1)\,\mathrm{Hz}$. From \eref{eq:598stark} we obtain
\begin{equation}
|\langle ^3D_1|| r || ^3P_1\rangle| = 1.265 (11)\,\, {\rm a.u.}
\end{equation}

\subsubsection{598 pole via scattering rate to stark shift ratio}
The 598 laser has the same polarization and frequency as used in the polarizability measurement.  The laser frequency is sufficiently close to the $\ket{^3D_1,7} \leftrightarrow \ket{^3P_1,6}$ transition that the scattering through $\ket{^3P_1,8}$ can be neglected. From the $\ket{^3D_1,7,0}$ state, the atom will scatter at the rate
\begin{equation}
\label{eq:598rate}
R_1 = \Gamma_1 \frac{4}{45} \frac{\Omega_1^2}{4 \Delta_1^2},
\end{equation}
where $\Gamma_1$ is the total decay rate of the $^3P_1$ state. From the ratio of \eref{eq:598stark} and \eref{eq:598rate}, $\Gamma_1$ is determined independent of $\Omega_1$.

The experimental procedure to measure $R_1$ is similar to the 646 case:
\begin{enumerate}
\item Repeat optical pumping into $\ket{^3D_1,7,0}$ and shelving to $\ket{^1S_0,7,-1}$ on the clock transition until the atom is detected dark.
\item Shelve $\ket{^1S_0,7,-1}$ back to $\ket{^3D_1,7,0}$
\item Apply 598 laser for duration $\tau$
\item Detect remaining $^3D_1$ population.
\end{enumerate}

\begin{figure}
\includegraphics[width=\columnwidth]{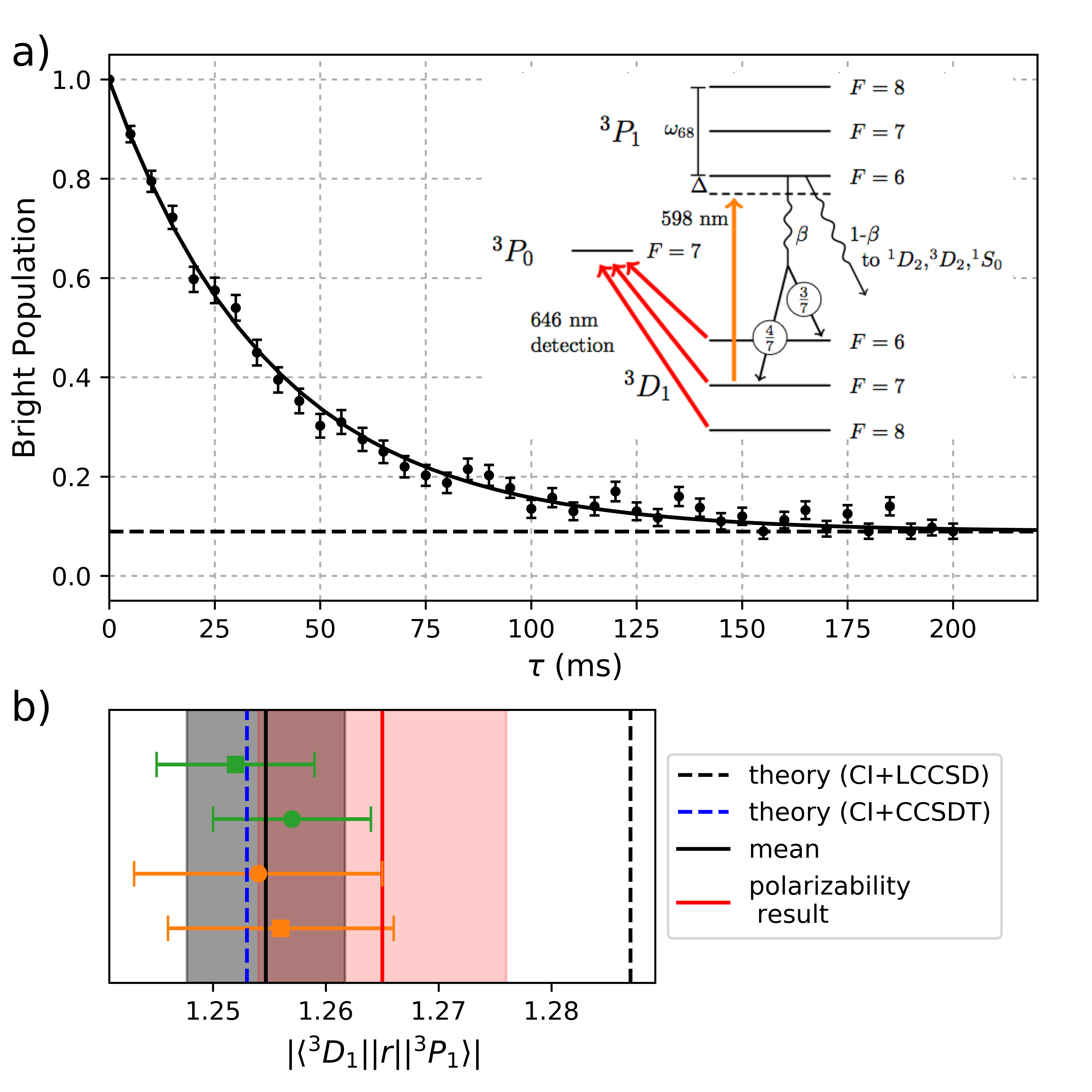}
\caption{(a) Bright population remaining after 598-nm laser pulse of duration $\tau$.  Solid black line is a fit to \eref{eq:598model}. a(inset) Schematic of 598-nm scattering experiment including branching paths from the $^3P_1$ $F$=$6$ state. (b) Comparison of results from different methodologies. Points are results from scattering to stark shift ratio experiment runs as described in the main text. Black line is the $\chi^2$ optimized mean of these results. Red line is the result from the polarizability measurement. Shaded regions indicate the uncertainties. The black dashed line is the theory value from previous work~\cite{luprop2016} and the blue dashed line is the new theory value from the method applied in this work, \sref{sec:theory}.  }
\label{fig:598scattering}
\end{figure}

The population dynamics are slightly complicated compared to the 646 case because ${^3D_1} \leftrightarrow {^3}P_1$ is an open transition.  The possible decay paths from $^3P_1$ $F$=$6$ are shown in \fref{fig:598scattering}a(inset), where $\beta$ is the branching ratio from $^3P_1\rightarrow {^3}D_1$. Scattering via $\ket{^3P_1,6,0}$ redistributes the populations, $p_6$ ($p_7$) in the $^3D_1$ $F$=$6~(7)$ hyperfine manifolds by the following rate equations:
\begin{align}
\frac{d p_6}{dt} &= \frac{3}{7} \beta R_1 p_7\\
\frac{d p_7}{dt} &= -\left(1-\frac{4}{7}\right) \beta R_1 p_7.
\end{align}
Solving for initial conditions $p_7(0) = P_0$ and $p_6 (0) = 0$, the bright population, $p_6$+$p_7$, after a pulse of length $\tau$ is
\begin{equation}
\label{eq:598model}
p(\tau) = P_0 \left[\frac{3\beta}{7-4\beta} +\frac{7-7\beta}{7-4\beta} e^{-R_1(1-\frac{4}{7}\beta)\tau} \right].
\end{equation}
The branching ratio $\beta$ was measured previously in $^{175}\mathrm{Lu}^+$ and reported to be $0.1862\,(17)$~\cite{luprop2016}. It has been remeasured in $^{176}\mathrm{Lu}^+$ (see Supplemental Material) and the same value was found with comparable uncertainty, $\beta = 0.1862\,(13)$.

The model, \eref{eq:598model}, does not account for the fact that population decaying to other magnetic substates in $^3D_1$ $F$=$7$ will subsequently scatter at different rates. However because the branching ratio back to $F$=$7$ is only $\frac{4}{7} \beta \approx 10 \% $ and the relative scattering rates for $\pi$-coupling from $|m| = (0,1,2) $ are close, (1.0,0.98,0.92), this is not expected to bias $R_1$  comparable to the reported uncertainty.

An example of the observed bright population after a 598-nm laser pulse of duration $\tau$ is shown in \fref{fig:598scattering}a. The dashed black line indicates the expected asymptotic bright population, $\frac{3\beta}{7-4\beta} = 0.08936\,(16)$, after the $^3D_1$ $F$=$7$ hyperfine manifold has been emptied. The solid black line is a fit to \eref{eq:598model} with $P_0$ and $R_1$ as free parameters. We acquire statistics on $P_0$, $R_1$, and $\delta_{1}$, with three interleaved experiments: (i) measure the bright population after preparation but without the 598-nm pulse to determine $P_0$ (ii) measure the population after a 598-nm pulse of fixed duration to determine $R_1$ from \eref{eq:598model}, and (iii) measure the ac-Stark on the $\ket{^3D_1,7,0}$ state using the 848-nm clock transition. A representative data set and statistical analysis is given in the Supplemental Material. From $P_0$, $R_1$, and $\delta_{1}$ the decay rate $\Gamma_1$ is found, which is related to the matrix element:
\begin{equation}
\beta \Gamma_1 = \frac{\omega_0^3 e^2 a_0^2}{3 \pi \epsilon_0 \hbar c^3} \frac{1}{2J^\prime+1}|\langle ^3D_1|| r || ^3P_1\rangle|^2,
\end{equation}
where $J^\prime$ is the total angular momentum of the excited state.

\fref{fig:598scattering}b shows the results from multiple experiment runs. Orange points were taken at detuning of $-1097.0(1)\,\mathrm{MHz}$ with respect to the $\ket{^3D_1,7,0} \leftrightarrow \ket{^3P_1,6,0}$ transition, while green points were at $+995.7(1)\,\mathrm{MHz}$ with respect to the $\ket{^3D_1,7,0} \leftrightarrow \ket{^3P_1,8,0}$ transition. The analysis is modified accordingly for scattering via $^3P_1$ $F$=$8$ (Supplemental Material). Square points used the conditional state preparation step as described, which prepares $P_0 \approx 99\%$ population in $\ket{^3D_1,7,0}$. Circles used only 646-nm optical pumping for state preparation which prepares $\approx 95\%$ of the population in $\ket{^3D_1,7,0}$, and $\approx 2\%$ each in $\ket{^3D_1,7,\pm1}$. From a full rate equation simulation including all sublevels, we find less than 0.15\% deviation of the scattering rate as compared to the model, \eref{eq:598model}, for either state preparation method.

The weighted mean result from the four experiment runs is
\begin{equation}
|\langle ^3D_1|| r || ^3P_1\rangle| = 1.255 (7)\,\, {\rm a.u.},
\end{equation}
indicated by the black line in \fref{fig:598scattering}b. The uncertainty in the mean is limited by the accuracy of $\beta$. The result is in agreement with the polarizability result (red in \fref{fig:598scattering}b).
\subsection{Polarizability measurement results and comparison with theory \label{sec:theory}}
For the remaining wavelengths, the differential dynamic scalar polarizabilities $\Delta\alpha_0(\omega)$ and tensor polarizabilities $\alpha_2(^3D_1,\omega)$ of the $^1S_0 \leftrightarrow {^3D_1}$ clock transition are found by the same methodology used in Ref.~\cite{kja2018static}.  By rotating an externally applied magnetic field to form angle $\phi\approx54.7^\circ$ with respect to the laser polarization, the ac Stark shift on the microwave transition vanishes and $\Delta\alpha_0(\omega)$ is inferred from the shift of the optical transition at the position of peak laser intensity. By setting $\phi=90^\circ$, the tensor polarizabilities $\alpha_2(^3D_1,\omega)$ are determined from the shift on the microwave transition. The measured ac-Stark shifts are given in the Supplemental Material and the inferred polarizabilities are summarized in \tref{tab:polresults}. The measurement wavelengths are given to an accuracy of 0.01 nm in \tref{tab:polresults} but are known more accurately (see Supplemental Material). A breakdown of the uncertainty budget for $\Delta \alpha_0$ at each wavelength is given in \tref{tab:poluncert}.

We compare polarizability measurements with theoretical calculations. In Ref.~\cite{luprop2016}, we used a method that 
combines configuration interaction (CI) and the linearized coupled-cluster single-double (LCCSD) approaches to study Lu$^+$. The application of this method to the calculation of polarizabilities was described in detail. 
In this work we further develop this method, additionally including  quadratic non-linear terms and (partially) triple excitations in the framework of the coupled-cluster approach to improve the effective Hamiltonian used in the CI calculation. The triple excitations are allowed from the core shells with principal quantum numbers $n=4,5$ to the virtual orbitals with maximal quantum numbers $n=15$ and $l=3$. Following the formalism developed in~\cite{PorDer06} we solve equations for triple cluster amplitudes iteratively, i.e., triples are included in all orders of the perturbation theory. The results obtained in the approach combining
CI and coupled-cluster single-double-triple (CCSDT) method (we refer to it as the CI+CCSDT method) are listed in~\tref{tab:polresults}.  At $\lambda = 10600$ nm, the theory is unable to provide a reliable prediction because the value is consistent with zero with the theoretical uncertainty. 
We use the effective (``dressed'') electric dipole operator in the polarizability calculations, which includes the random-phase approximation, core-Bruekner, two particle, structural radiation, and normalization corrections. A detailed description of these corrections is given in Ref.~\cite{DzuKozPor98}. The assignment of theoretical uncertainties is as discussed in Ref.~\cite{luprop2016}. As seen from the table, there is a good agreement between theory and experiment, though the experimental accuracy is better.

\begin{table}
\caption{Measured and calculated differential dynamic scalar polarizabilities $\Delta\alpha_0(\omega)$ and tensor polarizabilities $\alpha_2(^3\!D_1,\omega)$ (in a.u.).
The uncertainties are given in parentheses.}
\label{tab:polresults}
\begin{ruledtabular}
\begin{tabular}{l c c c c}
\multicolumn{1}{l}{} & \multicolumn{2}{c}{Experiment} & \multicolumn{2}{c}{Theory} \\
\multicolumn{1}{l}{$\lambda$ (nm)} & \multicolumn{1}{c}{$\Delta\alpha_0$} & \multicolumn{1}{c}{$\alpha_2$}& \multicolumn{1}{c}{$\Delta\alpha_0$} & \multicolumn{1}{c}{$\alpha_2$} \\
\hline \\ [-0.3pc]
  804.13  & $18.4\,(4)$           & $-13.97\,(31)$       & $22\,(4)$      & $-15.5\,(1.2)$ \\
  847.74  & $14.06\,(31)$         & $-11.59\,(26)$       & $17.2\,(3.9)$  & $-12.7\,(1.0)$ \\
  987.09  & $ 7.56\,(15)$         & $ -8.05\,(16)$       & $ 9.9\,(3.5)$  & $ -8.8\,(7)$ \\
 1560.80  & $ 2.22\,(6)$          & $ -5.73\,(15)$       & $ 3.6\,(3.1)$  & $ -5.9\,(5)$ \\
10600     & $0.059\, (4)^{\rm a}$ & $ -4.4\,(3)^{\rm a}$ & 1.2\,(2.9)     & $ -4.9\,(4)$
\end{tabular}

\end{ruledtabular}
\raggedright $^{\rm a}$These values were obtained in Ref.~\cite{kja2018static}.
\end{table}
 \begin{table}
\caption{Contributions to the uncertainty in $\Delta \alpha_0$ for each wavelength.}
\label{tab:poluncert}
\begin{ruledtabular}
 \begin{tabular}{l c c c c}
                                              &804        &848            &987           &1560 \\
effect  & \%& \%& \%& \%\\

\hline
beam profiling                            &0.9          &2.0		   &1.8	   &1.4\\
power measurement, statistical  &1.8          &0.6           &0.4          &2.3  	\\
power meter calibration 	         &0.8          &0.8           &0.8          &0.8\\
ac Stark shift, statistical            &0.06	  &0.06	  &0.1          &0.03	\\
\hline
total uncertainty                      &2.2	    &2.2	  	&2.0		  &2.7	\\
 \end{tabular}
\end{ruledtabular}
\end{table}

In \tref{E1_ME} we compare the absolute values of the reduced matrix elements $\langle 5d6s\,\,^3\!D_1 ||r|| 6s6p\,\,^3\!P_{0,1} \rangle$
obtained in this work with current experimental results and the values obtained in the framework of the CI+LCCSD approximation in Ref.~\cite{luprop2016}.
\begin{table}
\caption{\label{E1_ME} The absolute values of the reduced matrix elements
$\langle 5d6s\,\,^3\!D_1 ||r|| 6s6p\,\,^3\!P_{0,1} \rangle$ (in a.u.)
obtained in the CI+CCSDT method are compared them with the results obtained in the CI+LCCSD approximation in Ref.~\cite{luprop2016} and present experimental results. The uncertainties are given in parentheses.}
\begin{ruledtabular}
\begin{tabular}{lccccd}
\multicolumn{1}{l}{} & \multicolumn{1}{c}{Ref.~\cite{luprop2016}} & \multicolumn{2}{c}{This work} \\
\multicolumn{1}{l}{} & \multicolumn{1}{c}{CI+LCCSD} & \multicolumn{1}{c}{CI+CCSDT} & \multicolumn{1}{c}{Expt.} \\
\hline \\ [-0.3pc]
$|\langle ^3\!D_1 ||r|| ^3\!P_0 \rangle|$ & 1.480 & 1.430  & 1.440(2)  \\[0.2pc]

$|\langle ^3\!D_1 ||r|| ^3\!P_1 \rangle|$ & 1.287 & 1.253  & 1.255(7)  \\
\end{tabular}
\end{ruledtabular}
\end{table}
We find that the inclusion of the non-linear and triple terms into consideration significantly improved the agreement between the theoretical and experimental values.

\section{Modeling the differential polarizability}
The scalar dynamic polarizability of a given clock state $\ket{\nu}$ can be written as a positive sum of second order poles. In atomic units, this is given by~\cite{arora2007magic,polarizability2013}:
\begin{equation}
\alpha_0 (\omega) = \frac{2}{3(2J_\nu +1)} \sum_{\xi} \frac{\langle \xi ||r||\nu\rangle^2}{\omega_{\xi\nu}} \frac{1}{1-\left(\omega/\omega_{\xi\nu}\right)^2},
\end{equation}
where $\langle \xi ||r||\nu\rangle$ is the reduced dipole matrix element for transition at frequency $\omega_{\xi\nu}=E_\xi-E_\nu$, and $J_\nu$ is the total angular momentum of state $\ket{\nu}$.

Using the identity
\begin{equation}
\label{EqResidual}
\frac{1}{1-x^2} = \frac{x^{2(n+1)}}{1-x^2}+\sum_{k=0}^n x^{2k}
\end{equation}
any pole can be split into the sum of a polynomial of order $2n$ and a term that is henceforth referred to as the pole residual.  From calculated matrix elements \cite{luprop2016}, pole residuals for each contributing transition can be calculated at each measurement wavelength. For $n=2$, these results are tabulated in table~\ref{residuals} along with subtotals for each clock state. The two dominant contributions from the 598 and 646-nm transitions are omitted from the $^3D_1$ subtotal.  As the residuals are less significant at longer wavelengths, only results for  the measurement wavelengths of 804 and 848\,nm are given.
\begin{table}
\caption{\label{residuals} The $n=2$ residuals for each pole contributing to $\Delta\alpha_0(\omega)$ evaluated at $\omega_{804}$ and $\omega_{848}$.  For the 598- and 646-nm poles, the values are determined from the measured matrix elements.  All others are taken from theory~\cite{luprop2016} using experimental energies.  Subtotals given for the $^3D_1$ state omit the two dominant contributions from the 598- and 646-nm poles.}
\begin{ruledtabular}
\begin{tabular}{llcc}    
\multicolumn{1}{l}{State}
&\multicolumn{1}{l}{Contribution} & \multicolumn{1}{c}{804-nm}
& \multicolumn{1}{c}{848-nm} \\
\hline \\ [-0.3pc]
$6s^2\;^1\!S_0$ &$6s6p\;^3\!P_1^o$& 0.029 & 0.021  \\
                &$6s6p\;^1\!P_1^o$           & 0.063 & 0.045  \\
                &$5d6p\;^3\!D_1^o$           & 9.4[-4] & 6.9[-4]  \\
                &$5d6p\;^3\!P_1^o$           & 1.5[-4] & 1.1[-4]  \\
                &$5d6p\;^1\!P_1^o$           & 4.1[-4] & 3.0[-4]  \\
                \\
                & Total                                                         & 0.094 & 0.067 \\[0.5pc]

$5d6s\;^3\!D_1$ &$6s6p\;^3\!P_0^o$& 4.990 & 3.074 \\
                &$5d6p\;^3\!P_0^o$           & 6.0[-3] & 4.4[-3] \\[0.3pc]

                &$6s6p\;^3\!P_1^o$           & 1.752 & 1.135 \\
                &$5d6p\;^3\!D_1^o$          & 0.023 & 0.017 \\
                &$5d6p\;^3\!P_1^o$           & 7.3[-3] & 5.3[-3] \\[0.3pc]

                &$6s6p\;^3\!P_2^o$           & 0.022 & 0.015 \\
                &$5d6p\;^3\!F_2^o$           & 0.086 & 0.061 \\
                &$5d6p\;^1\!D_2^o$           & 0.012 & 9.0[-3]  \\
                &$5d6p\;^3\!D_2^o$          & 0.016 & 0.011  \\
                &$5d6p\;^3\!P_2^o$          & 4.2[-4] & 3.0[-4]  \\
                \\
                & Subtotal                                                   & 0.174 & 0.124 \\
\end{tabular}
\end{ruledtabular}
\end{table}

As seen from the table, the residual contribution from either clock state is at most the measurement error for any given measurement wavelength and, even then, there is a significant cancellation between them.  The omission of these residuals is then well justified even for rather significant changes to the theoretical calculations.  Additionally, with $1\%$ accuracy on the contributions from 598 and 646, the error from this is no more than $30\%$ of the measurement error.  Hence, $\Delta\alpha_0(\omega)$ can be modeled by
\begin{multline}
\label{EqModel1A}
\Delta \alpha_0(\omega)=\frac{2}{9}\frac{\mu_{598}^2}{\omega_{598}}\frac{1}{1-\left(\omega/\omega_{598}\right)^2}+\frac{2}{9}\frac{\mu_{646}^2}{\omega_{646}}\frac{1}{1-\left(\omega/\omega_{646}\right)^2}\\
+a_0+a_1 \left(\frac{\omega}{\omega_{804}}\right)^2+a_2 \left(\frac{\omega}{\omega_{804}}\right)^4,
\end{multline}
where $\mu_\lambda$ are the reduced electric dipole matrix elements for the respective transitions, $\omega_{646}$ and $\omega_{598}$ are the respective resonant transition frequencies, and $a_k$ are polynomial fitting coefficients. The scaling of the frequency for the polynomial terms is arbitrary and conveniently set to the largest frequency in the measurement window, $\omega_{804}$.  Using \eref{EqResidual} and a suitably modified $a_k$, this can be rewritten in the mathematically equivalent form
\begin{equation}
\Delta \alpha_0=b_\mathrm{598}(\omega)+b_\mathrm{646}(\omega)+\sum_{k=0}^{2} a_k \bar{\omega}^{2k},
\end{equation}
where $\bar{\omega}=\omega/\omega_{804}$ and
\begin{equation}
b_\lambda(\omega)=\frac{2}{9}\frac{\mu_\lambda^2}{\omega_\lambda}\frac{\left(\omega/\omega_\lambda\right)^6}{1-\left(\omega/\omega_\lambda\right)^2}.
\end{equation}
Values for $a_k$ can then be found from a $\chi^2$-minimization.

As the fitting function is a linear combination of bases functions, the minimization can be elegantly solved using singular-valued decomposition (SVD).  The functional form of $b_\lambda(\omega)$ is practically exact as the transition frequencies are well known \cite{dean2018}.  Only the overall scale, which is determined by the relevant matrix element (squared), is subject to experimental uncertainty.  For now we assume these are exact.  With measurements, $\vect{m}_\mathrm{j}$, of the polarizability performed at $\omega_j$ with uncertainties $\sigma_j$, we seek to find coefficients $\vect{a}$ via the $\chi^2$-minimization,
\begin{equation}
\min_\vect{a} \lVert \vect{A}\cdot\vect{a}-\vect{b} \rVert^2
\end{equation}
where
\begin{equation}
(\vect{A})_\mathrm{jk}=\frac{\bar{\omega}_j^{2k}}{\sigma_j}, \quad \vect{b}_\mathrm{j}=\frac{(\vect{m}-\vect{b}_{598}-\vect{b}_{646})_\mathrm{j}}{\sigma_j},
\end{equation}
and $(\vect{b}_{\lambda})_\mathrm{j}=b_{\lambda}(\omega_j)$.  With the SVD, $\vect{A}=\vect{U}\vect{S}\vect{V}^T$, the solution is then
\begin{equation}
\vect{a}=\vect{V}\vect{S}^{-1}\vect{U}^T\vect{b},
\end{equation}
where $\vect{S}^{-1}$ is to be interpreted as the left inverse.  The polarizability at any given frequency is then given by
\begin{equation}
\label{ExpPol}
\Delta\alpha_0(\omega)=\vect{a}\cdot\vect{v}\left(\bar{\omega}\right)+b_{598}(\omega)+b_{646}(\omega)
\end{equation}
with $\vect{v}(x)=(1,x^2,x^4)$.

In terms of errors there are two distinct considerations.  The first is simply the error associated with the fit, which arises from the first term in the equation for $\alpha_0(\omega)$.  As this is a linear combination of the coefficients $\vect{a}$, the 1-$\sigma$ error is given by
\begin{equation}
\label{fiterror}
\delta \alpha_0(\omega)=\sqrt{\vect{v}(\bar{\omega})^T (\vect{A}^T\vect{A})^{-1} \vect{v}(\bar{\omega})},
\end{equation}
which cannot be treated as uncertainties in each of the polynomial coefficients.  Although each coefficient $a_k$ can be prescribed an uncertainty, each of these uncertainties has some degree of correlation which is accounted for by Eq.~\ref{fiterror} insofar as the evaluation of the polarizability at a given frequency is concerned.

The second consideration is from an error in $b_\lambda(\omega)$.  Varying either of these by the fractional amount $\sigma_\lambda$ will change the solution by
\begin{multline}
\label{poleerror}
\delta \alpha_0(\omega)=-\left(\vect{v}(\bar{\omega})^T\vect{V}\vect{S}^{-1}\vect{U}^T\left(\frac{\vect{b}_\lambda}{{\bm \sigma}}\right)\right)\sigma_\lambda\\
+b_\lambda(\omega)\sigma_\lambda,
\end{multline}
where ${\bm \sigma}$ is the vector of measurement uncertainties, $\sigma_j$, and the vector division is be interpreted element-wise. As $\vect{v}$ and $\vect{A}$ are unaffected, the error given by Eq.~\ref{fiterror} is unchanged. Note that both $\sigma_{598}$ and $\sigma_{646}$ are $\lesssim 0.01$.  As it turns out, these errors are much smaller than those from the fit and can be largely ignored being no more than around $5\%$ of the fitting error over the frequency range of interest.  The reason for this is that small changes in $b_\lambda(\omega)$, that may make significant changes to the polarizability, are largely compensated by the fitting so as to remain consistent with the measurements.

\begin{figure}
\includegraphics[width=\columnwidth]{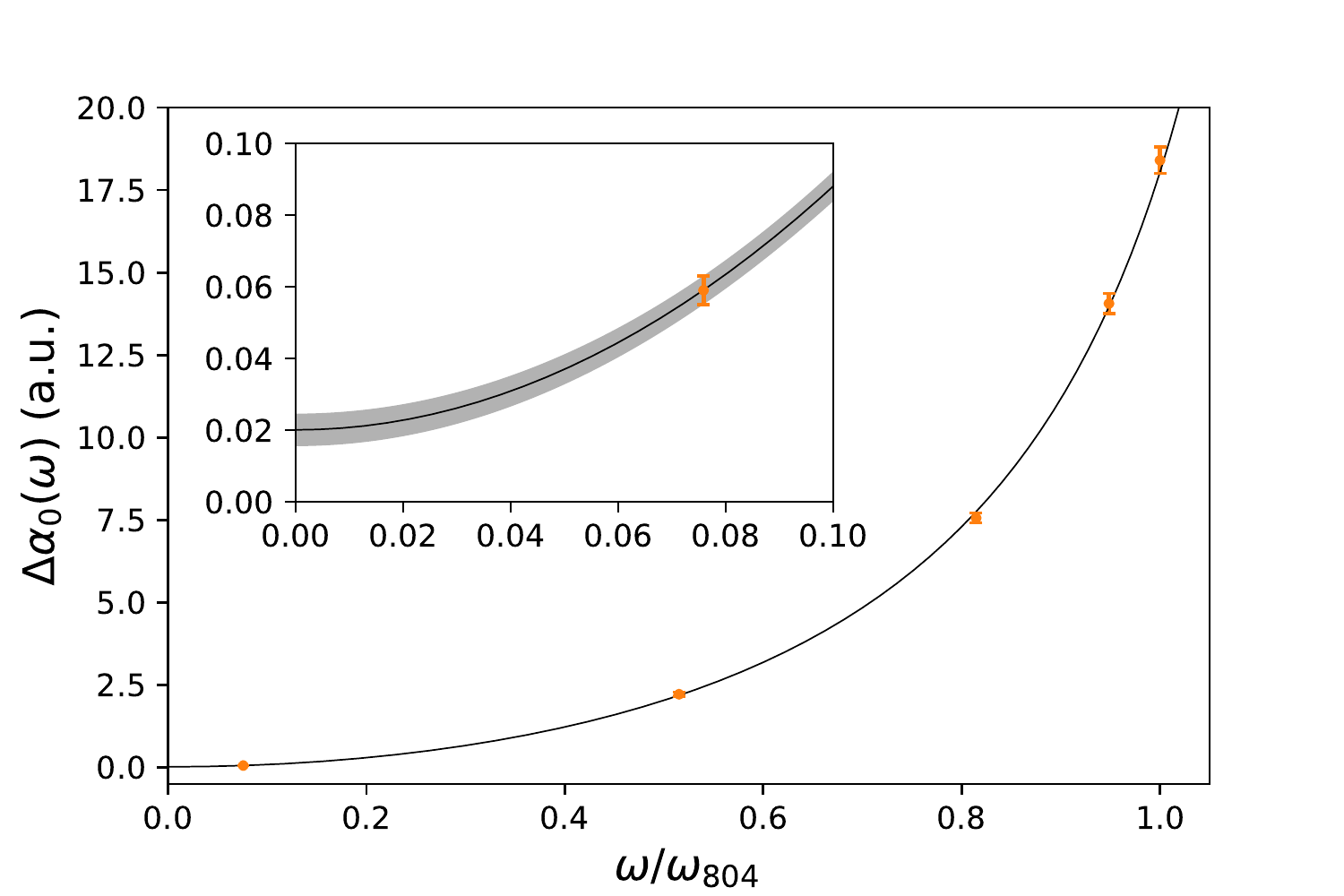}
\caption{Polarizability measurement results (orange points) from \tref{tab:polresults} and fit to the model given by \eref{ExpPol}  (solid black line). The inset shows the model uncertainty (gray shaded) near dc, which is predominately determined by the measurement uncertainty at 10.6 $\mu$m (orange point inset).}
\label{fig:polresults}
\end{figure}

The result of the minimization procedure described above is shown in \fref{fig:polresults}. We find

\begin{equation}
\Delta\alpha_0(0)=0.0201(45)
\end{equation}
with a reduced $\chi^2=1.48$.  The extrapolated value is consistent with that determined from the measurement at $10.6\,\mathrm{\mu m}$ and extrapolated using theory \cite{kja2018static}.  The error bar also reflects the intuitively obvious fact that the error in the extrapolation cannot be better than the measurement error at $10.6\,\mathrm{\mu m}$: with a three parameter fit to five data points there is insufficient averaging to expect better particularly with the other measurements far removed from the extrapolation point.  This should be contrasted with the claim in \cite{YbPeik2016}.

It is of interest to compare and contrast the model used with the single pole approximation~\cite{YbPeik2016}
\begin{equation}
\Delta\alpha_0(\omega)\approx c_0+\frac{c_1(\omega/\omega_0)^2}{1-(\omega/\omega_0)^2},
\end{equation}
which may be viewed as a Pad{\'e} approximant accurate to 4\textsuperscript{th} order.  In general, care should be taken with such an approximation: it constrains the relative signs of the quadratic and quartic terms and this need not be the case for a differential polarizability in which there can be significant pole cancelation.  In the case of lutetium, $\Delta\alpha_0(\omega)$ is dominated by two transitions connected to $^3D_1$ which insures the relative sign.  Moreover the two poles are closely spaced and reasonably removed from the measurement window of interest.  Hence such an approximation may be reasonable.  Fitting to this model gives $\Delta\alpha_0(0)=0.0203(42)$  with a reduced $\chi^2$ of 0.94, in complete agreement with the previous fit.  Additionally, the effective pole at $\omega_0$ has a wavelength of $639(7)\,\mathrm{nm}$ consistent with the expectation that it lies between the two dominant poles at 598 and 646\,nm and weighted towards the strongest contribution at $646\,\mathrm{nm}$.  As there is no significant modeling dependence, we use the more general model in assessing the BBR shift for the convenience that comes with the linear parameter dependence.

\section{The BBR shift assessment}
With $\Delta\alpha_0(\omega)$ experimentally characterized up to $\omega_{804}$, the BBR shift can be readily calculated.  Integrating over the BBR spectrum we have
\begin{align}
\delta\nu&=-\frac{1}{2} \int_0^\infty \Delta\alpha_0(\nu) \frac{8\pi h}{c^3\epsilon_0}\frac{\nu^3}{e^{h\nu/(k_B T)}-1}\,\mathrm{d}\nu\\
&=-\frac{1}{2} \left(\frac{8\pi^5 k_B^4 T_0^4}{15 h^3 c^3\epsilon_0}\right) \left(\frac{T}{T_0}\right)^4 \nonumber\\
&\quad\quad \times \frac{15}{\pi^4} \int_0^\infty \Delta\alpha_0(u) \frac{u^3}{e^{u}-1}\,\mathrm{d}u\\
&=-\frac{1}{2} \left(831.945 \,\mathrm{V/m}\right)^2 \left(\frac{T}{T_0}\right)^4 \nonumber\\
&\quad\quad \times \frac{15}{\pi^4} \int_0^\infty \Delta\alpha_0(u) \frac{u^3}{e^{u}-1}\,\mathrm{d}u,
\end{align}
where $T_0=300\,\mathrm{K}$ and $u=h\nu/(k_B T)$ is a dimensionless scale of integration.  Defining $\bar{T}=T/T_0$ and using Eq.~\ref{ExpPol}, the BBR shift can be written
\begin{equation}
\delta\nu=-\frac{1}{2} \left(831.945 \,\mathrm{V/m}\right)^2 \bar{T}^4 \mathbf{a}\cdot\mathbf{w}(\bar{T}),
\end{equation}
where
\begin{equation}
\vect{w}(\bar{T})=\left(1,\frac{40\pi^2}{21} \epsilon^2 \bar{T}^2,8\pi^4\epsilon^4\bar{T}^4\right),\quad \epsilon=\frac{k_B T_0}{\hbar\omega_{804}}.
\end{equation}
From the fitted coefficients, the fractional BBR shift is then given by
\begin{equation}
\frac{\delta \nu}{\nu}=-4.90\times 10^{-19}\bar{T}^4 (1+1.77 \bar{T}^2),
\end{equation}
where the term proportional to $\bar{T}^8$ has been omitted as it contributes only $\sim1\%$ at 300$\,$K.


 As with the polarizability itself, only the fitting error significantly influences the uncertainty.  Also, although the BBR shift is best given as an expansion of varying powers of temperature, the uncertainty in its estimate is best represented by a term similar to Eq.~\ref{fiterror} and not independent uncertainties of the expansion coefficients. Explicitly
\begin{equation}
2.45\times 10^{-17}\bar{T}^4 \sqrt{\vect{w}(\bar{T})^T (\vect{A}^T\vect{A})^{-1} \vect{w}(\bar{T})},
\end{equation}
and, over the practical temperature range of 270-330\,K, this is well approximated by $9.8\times 10^{-20} \bar{T}^4.$  Corrections due to Eq.~\ref{poleerror} are less than $1\%$ of this expression. The BBR shift at room temperature is then  $-1.364(98)\times 10^{-18}$ in agreement with the previous assessment \cite{kja2018static}.

\section{Conclusion}
In summary, the differential polarizability of the $^{176}$Lu$^+$ $^1S_0\leftrightarrow {^3}D_1$ clock transition has been measured over a range of wavelengths.  This has allowed an extrapolation to the true static value relevant to micromotion clock shifts and an experimental determination of the dynamic correction to the BBR shift.  Model dependency for the extrapolation was investigated using two independent fitting models: both of which could be justified based on theoretical considerations and gave excellent agreement in the extrapolated value.

The experimental determination of intensities is a crucial component of the polarizability assessment and this was rigorously tested using an independent polarizability measurement near to resonance with two contributing transitions.  We consider this an essential consistency check when using extrapolation of high accuracy polarizability measurements for BBR shift assessments. Such a consistency check is readily available for any clock candidate having a transition associated with detection. The measurements also provided precision benchmarks for the theoretical approach developed in this work.

\section{Acknowledgements}
This work is supported by the National Research Foundation, Prime Ministers Office, Singapore and the Ministry of Education, Singapore under the Research Centres of Excellence programme. It is also supported by A*STAR SERC 2015 Public Sector Research Funding (PSF) Grant (SERC Project No: 1521200080). T. R. Tan acknowledges support from the Lee Kuan Yew postdoctoral fellowship.
 This work was supported in part by the Office of
Naval Research, USA, under award number N00014-17-1-2252 and Russian Foundation for Basic Research under
Grant No. 17-02-00216.


%

\pagebreak



\section*{Supplementary material}

# Supplemental Material for Dynamic polarizability measurements in $^{176}$Lu$^+$

K. J. Arnold[1],[*] R. Kaewuam[1], T. R. Tan[1,2], S. G. Porsev[3,4], M. S. Safronova[3,5], and M. D. Barrett[1,2][†]

[1]*Centre for Quantum Technologies, 3 Science Drive 2, 117543 Singapore*
[2]*Department of Physics, National University of Singapore, 2 Science Drive 3, 117551 Singapore*
[3]*Department of Physics and Astronomy, University of Delaware, Newark, Delaware 19716, USA*
[4]*Petersburg Nuclear Physics Institute of NRC "Kurchatov Institute", Gatchina, Leningrad District 188300, Russia*
[5]*Joint Quantum Institute, National Institute of Standards and Technology and the University of Maryland, College Park, Maryland 20742, USA*

Supporting data for power meter calibration and intensity characterization including detailed uncertainty assessments. Technical details on the lasers sources used. Measurement of the $^3P_1 \to {}^3D_1$ branching ratio. Supporting data for the 646 nm and 598 nm scattering to Stark shift ratio measurements, and all polarizability measurements.

PACS numbers: 06.30.Ft, 06.20.fb

## CONTENTS

## I. POWER MEASUREMENT AND CALIBRATION

Polarizability measurements were performed over the course of approximately five months using the following detectors, given with their manufacturer specifications:

| Label | Manufacturer | Model | Type | λ Range (nm) | Power Range | Linearity | Accuracy |
|---|---|---|---|---|---|---|---|
| A | Thorlabs | S121C | Si | 400-1100 | 500 nW - 500 mW | ±0.5% | ±3% |
| B | Thorlabs | S121C | | | | | |
| C | Thorlabs | S121C | | | | | |
| D | Thorlabs | S120VC | Si | 200-1100 | 50 nW - 50 mW | ±0.5% | ±3% |
| E | Newport | 918-SL | Si | 400-1100 | 20 pW-2.5 mW | ±0.5% | ±1% |
| F | Thorlabs | S122C | Ge | 700-1800 | 50 nW - 40 mW | ±0.5% | ±5% |

[*] cqtkja@nus.edu.sg
[†] phybmd@nus.edu.sg

After completion of the measurements, detectors 'A' and 'F' were calibrated by Singapore's National Metrology Centre (NMC). Calibration was performed against one of NMC's 'working standards' which are maintained by biennial recalibration against their cryogenic radiometer primary standard. Calibration was performed at the closest wavelengths to our measurement wavelengths that NMC provides, which was limited to their available laser sources. The results of the calibrations are:

| Detector | Calibration Wavelength (nm) | Deviation | Uncertainty ($1\sigma$) |
|---|---|---|---|
| A | 850 | +2.4% | ±0.75% |
| A | 633 | +2.9% | ±0.75% |
| F | 976 | -1.4% | ±0.75% |
| F | 1550 | -0.23% | ±0.75% |

Polarizability measurements at 986 nm and 1560 nm both used detector 'F' as the reference detector and the power readings were corrected based on calibrations at 976 nm and 1550 nm, respectively. Detector 'A' was not exclusively used as the reference detector for all polarizability measurements at the other wavelengths (598, 646, 804, 848) and it was necessary to transfer the calibration of detector 'A' to other detectors. This was done in a test setup using actively power-stabilized laser sources with variable attenuation to compare detectors 'A' to 'E' over the dynamic range used in reported experiments. The transfer calibration measurements are given in Fig. 1.

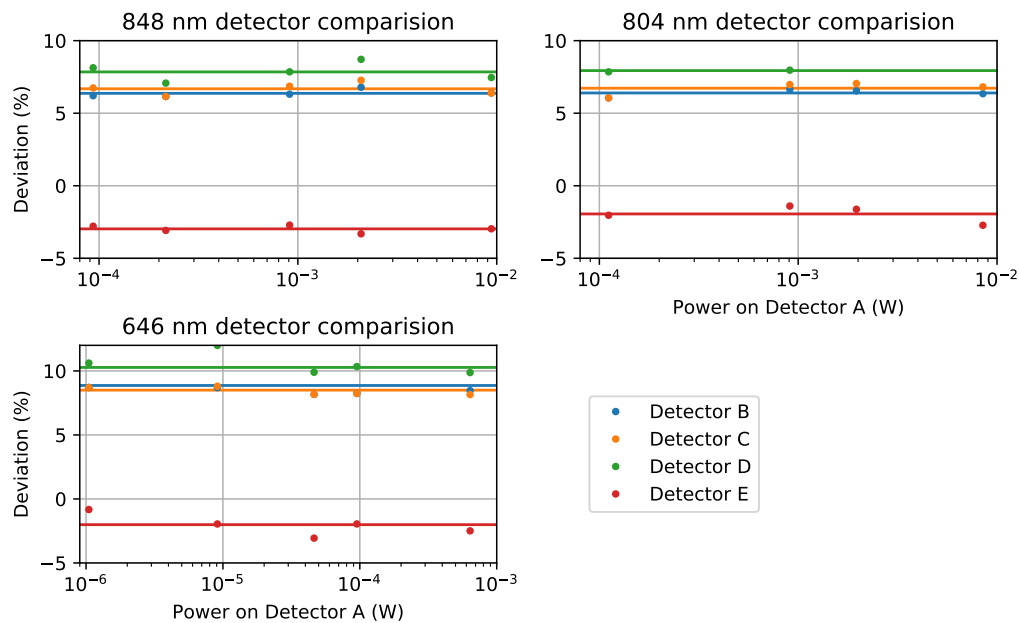

FIG. 1. Transfer calibration measurements at 848, 804, and 646 nm to assess accuracy of detectors B-E with respect to detector A, which was calibrated by NMC. Solid lines are the mean deviation for measurements over the entire power range explored.

The average deviation of each detector with respect to the NMC calibration, using detector 'A' as the transfer standard, are summarized below. In parenthesis is the standard deviation of the measurements over the range of powers explored. We note the assessed calibration errors are $2\sigma$ or more larger than the manufacturer specified accuracy for every detector B through E.

| Detector | 848 | 804 | 646 |
|---|---|---|---|
| B | +6.4(0.2)% | +6.4(0.2)% | +8.9(0.9)% |
| C | +6.7(0.5)% | +6.7(0.4)% | +8.5(0.3)% |
| D | +7.8(0.6)% | +7.9(0.1)% | +10.3(0.9)% |
| E | -2.9(0.2)% | -1.9(0.5)% | -2.0(0.7) |

Table I lists the power measurements for the experiments at each laser wavelength, using the methodology described

in the main article (refer to Figure 1 in the article). The reference detector listed is corrected by the calibration factor $\eta$ linking each reference detector to an NMC calibration at the nearest wavelength. The NMC calibration at 633 nm is used for both 598 and 646 nm measurements. The NMC calibration at 850 nm is used for both 804 and 848 nm measurements. Here $T = (P_1 - P_2 - P_3)/P_1$ is the transmission of vacuum viewport. The vacuum viewport is broadband AR coated for $780-1560\,\mathrm{nm}$ at normal incidence, but still exhibits high transmission down to 646 nm and at the 30° angle of incidence used for the experiments. The ratio of the reference detector reading (at $P_1$) and monitor detector reading (at $P_4$ except for special cases described below) is $r = P_\mathrm{ref}/P_\mathrm{mon}$, measured while the power stabilization is necessarily disengaged. With the stabilization engaged, a reading is taken from the monitor detector $\tilde{P}_\mathrm{mon}$. The power at the ion is then $P_0 = \eta T r \tilde{P}_\mathrm{mon}$. For the 598 and 1560 nm cases there are minor differences in the implementation. For 598 nm, because of the low optical powers involved, the power stabilization photodiode was placed at position $P_4$ and also served as the monitor detector. For 1560 nm, the monitor detector was placed at $P_3$ because the laser power at $P_4$ would greatly exceed the maximum range of the detector head.

TABLE I. Summary optical power measurements.

| $\lambda$ | Reference Detector | $\eta$ | $T$ | $r$ | $\tilde{P}_\mathrm{mon}$ | $P_0$ |
|---|---|---|---|---|---|---|
| 598 | E | 1.020(8) | 0.9146(20) | 2.476(30) $\mu$W/V | 1.201(1) $V$ | 2.77(4) $\mu$W |
| 656 | A | 0.971(8) | 0.9818(3) | 0.863(5) | 44.1(1) $\mu$W | 48.7(5) $\mu$W |
| 804 | B | 0.936(8) | 0.9902(2) | 1.013(18) | 13.30(1) mW | 12.49(25) mW |
| 848 | B | 0.936(8) | 0.9860(1) | 1.039(7) | 16.88(1) mW | 16.19(17) mW |
| 986 | F | 1.014(8) | 0.9799(1) | 1.088(4) | 27.90(1) mW | 29.18(26) mW |
| 1560 | F | 1.002(8) | 0.99422(8) | 415(9) | 1.068(1) mW | 441 (10) mW |

## II. POWER STABILITY

When the active power stabilization is engaged, the reading on the monitor detector is stable to $\pm 1$ in the least significant display digit on the power meter console and repeatable from one day to the next. To further assess the power stability, we check the Allan deviation of the Stark shift as measured at the ion by the interleaved servo. Typical Allan deviations are shown in the Fig. II, taken from Stark measurements at 804 nm and 646 nm for example. This indicates the power instability with active stabilization is better than $10^{-3}$ and power stability is not significant source of uncertainty in the polarizability measurements.

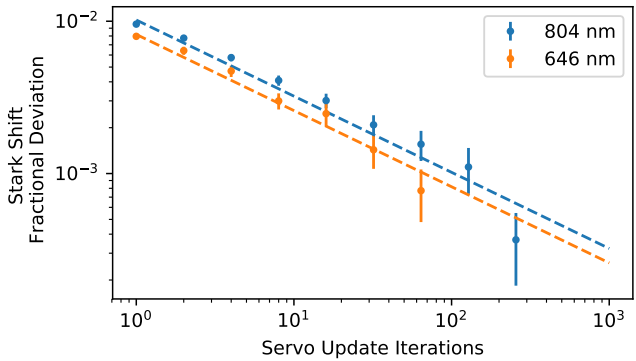

FIG. 2. Allan deviation of the Stark shift measured by the interleaved servo at the position of peak beam intensity. For the respective data sets the mean stark shifts are -340.2 (1) Hz and -846.5 (3) Hz, measurement durations are 90 minutes and 20 minutes, and the dashed lines indicate the projection noise limits.

## III. TRANSLATION STAGES AND POSITIONING ACCURACY

As shown in Figure 1 of the main article, a CCD camera detects one reflection of the Stark laser from a glass pick off. The beam position on the camera is determined by a 2D gaussian fit to the image. First, we consider the stability

of the beam position when the translation stages are static. Fig. IIIa shows the position determined from repeated position measurements on the camera every 5 seconds for one hour. The standard deviation is ±150 nm which we take as the measurement uncertainty in the beam position as determined by the camera. Fig. IIIb shows the position monitored on the camera over 34 hours. The position drifts by approximately 3 $\mu$m over the course of the day with maximum drift rate of 9 $\mu$m/day during the daytime and is relatively stable at night. The camera and stage are separated on the optical table by approximately 20 cm, and given a thermal expansion coefficient of 15 ppm/C° for stainless steel, we would roughly expect ∼3 $\mu$m/$C°$ variation of the beam position with lab temperature. The lab temperature typically increases by ∼1 C° during the daytime.

The camera and ion trap are also separated by about the same distance (20 cm) and so we must assume that the relative position of the camera and ion trap also drifts by a similar amount as observed in Fig. IIIb. The typical duration of a profile scan is 8 hours, usually started in the evening and run overnight. In Sec. IV we consider the effect of this thermal positioning drift in the uncertainty assessment for the beam profiles.

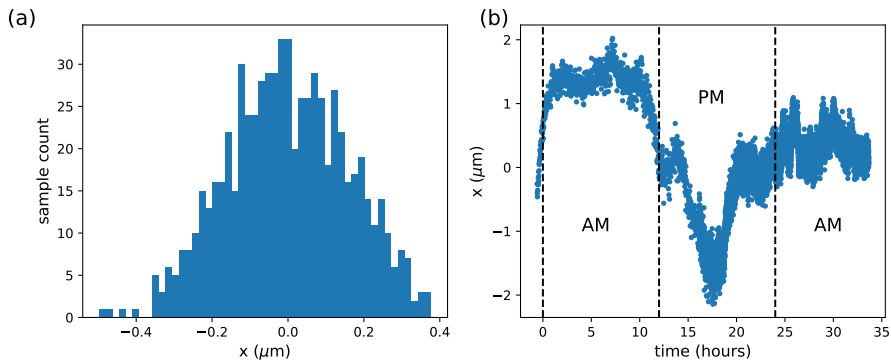

FIG. 3. (a) Histogram of repeated position measurements by the camera over one hour. (b) Beam pointing stability at the camera over 34 hours.

Figure IIIa shows a typical grid scan, in this case for one of the 646 nm profile scans. To minimize backlash of the motorized linear translation stages (Thorlabs Z725B), the scan is performed with forward actuator movements as much as possible. Starting from the point labeled '1', the $x$ stage is stepped forward across one row. Following advancement of the $y$ stage to the next row, the $x$ stage is retracted 100 $\mu$m past the next point ('20') and moved forward into position, and so forth. The blue points in Fig. IIIa indicate the set positions of the the stages, and the orange points are the displacement as measured by the CCD camera. Fig. IIIb shows the position errors in the $x$ and $y$ directions for each point. Here the rms error for the $y$ displacement is 0.6 $\mu$m and for the $x$ displacement 1.1 $\mu$m. By using the beam positions measured by the CCD camera for the analysis of the beam profile data instead of the stage set positions, the stage positioning errors do not effect the beam profile uncertainty. Only in the case of the 1560 nm profile, where it was not possible to use the CCD camera, are the programmed stage coordinates used for the subsequent analysis. For that case, the positioning errors as observed here are included in uncertainty assessment for the beam profile.

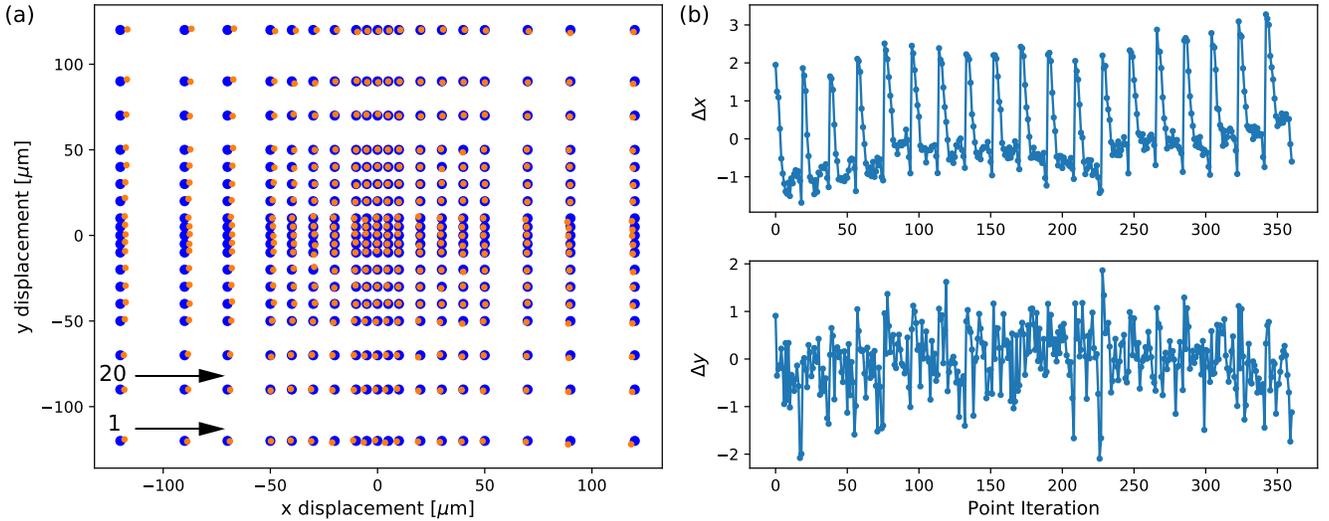

FIG. 4. (a) Representative 2D scan for beam profiling. Blue points are the set positions and orange points are the measured positions by the camera. (b) Position errors with respect to the programmed coordinates as measured by the camera at each position iteration.

## IV. BEAM PROFILE ANALYSIS

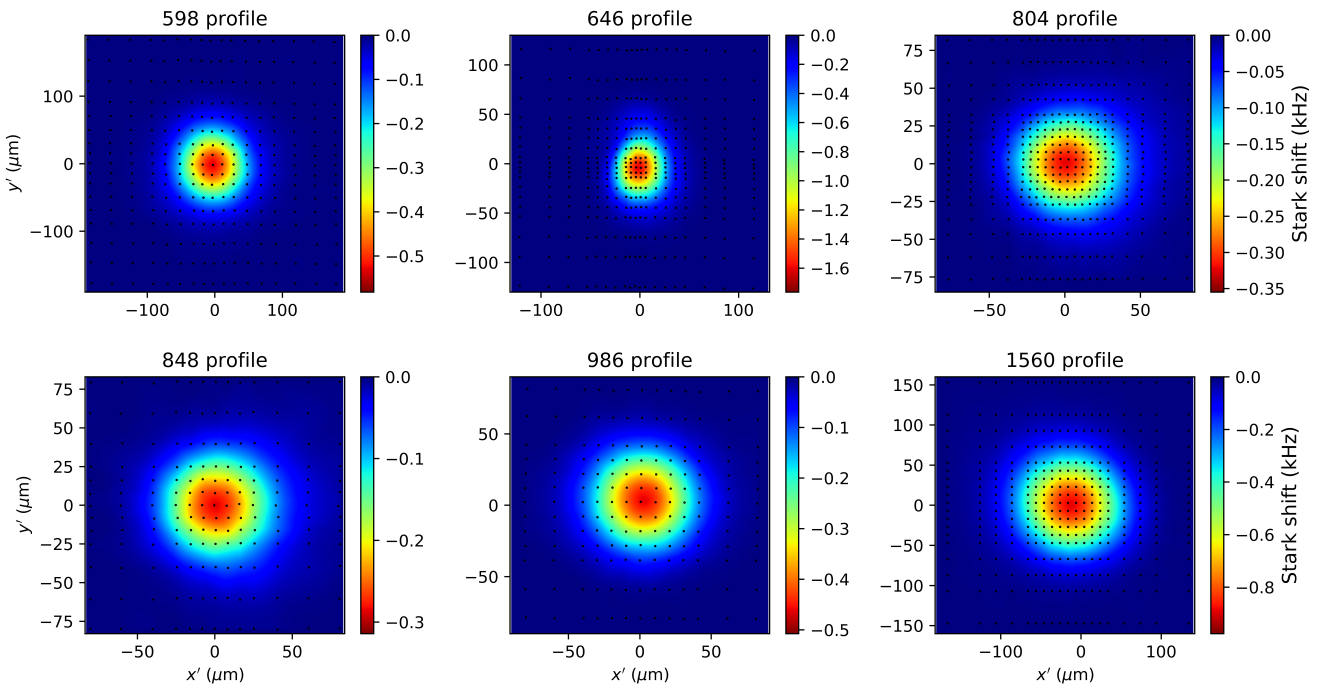

FIG. 5. Beam profiles at each wavelength. Colormaps are cubic splines to the Stark shift measurements taken at positions indicated by black dots. All beams have an astigmatism resulting from focusing through the glass pick-off window and the optical viewport at an angle (see setup in main article Fig. 1). This is most noticeable for 646 nm profile. For 646 nm, the beam was characterized with a camera at several positions near to the focal plane outside vacuum and the elliptically of the *in situ* 646 nm profile shown below is fully consistent with one of camera images which was 3 mm displaced from the focal plane where beam is circular.

Figure IV shows the beam profiles taken at each wavelength. The projection noise limited uncertainty per point is in the range of 1.0 to 1.5 Hz for all profiles. Measurement duration for the displayed profiles varies between 4.5 and 11 hours. The normalization constant $C$ is determined by the ratio of the peak stark shift divided by the numeric integral of a cubic spline to the data. A potential source of systematic error in determining $C$ is power distributed outside the measurement window. As discussed in the main article and demonstrated by example, low intensity distributed over a large area in the tails of the intensity distribution may be below the projection noise levels but still lead to a significant systematic error. This is assessed by profiling of the beam focus with a camera outside vacuum in a test setup using the same optics and an identical vacuum viewport. For 646 and 598 nm, the beam was profiled with a low noise Hamamatsu ORCA-Flash4.0 CMOS camera giving $\sim 10^{-4}$ signal to noise relative to the peak intensity. For 986 nm, the profile was captured with a Lumenera Inifinity3 camera, which provided only $\sim 10^{-3}$ signal to noise. Fig. IV shows the fraction of power captured when the integration region is restricted to a square with halfwidth $l$ scaled by the effective waist $w_e$ for the respective camera beam profiles. We note in all cases the convergence is similar and consistently slower than for a TEM$_{00}$ gaussian mode due to the fatter tails of the true intensity profiles. The colored dots indicate the size of the square grid used for the Stark shift profiles and the estimated power outside of the scan region based on their respective camera profiles. For 804 and 848 nm, the camera profiles (Lumenera Inifinity3) were taken only to obtain an approximate measure of the beam waist, at that time, and were unfortunately not of sufficient quality to extract useful quantitive information at a later date. However, since these wavelength used the same focusing assembly as for 986 nm, with only minor adjustment of the aspheric lens to position the focus at the ion, the 986 nm camera profile is used to estimate the power loss factor. For 1560 nm, we do not have a suitable camera to profile this wavelength. Given that the mode is conditioned in the same way, i.e. collimated from an optical fiber by an aspheric lens and focused with an acromat lens, we estimate the power capture factor based on the other available profiles in Fig. IV. In every case, $C$ is corrected for the estimated power capture, and an additional uncertainty included for the full magnitude of the correction (see Table II).

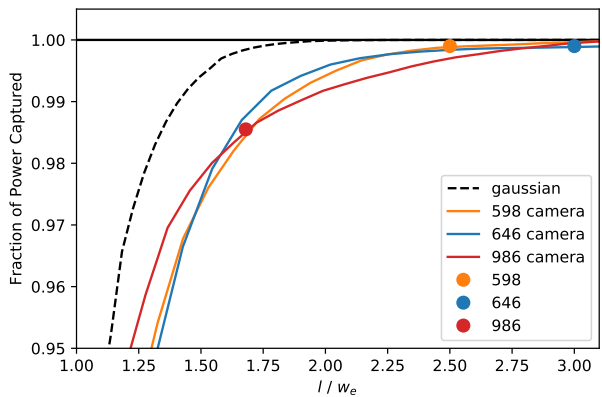

FIG. 6. Solid lines are the power capture within a square region of halfwidth $l$ scaled by the effective waist $w_e$ for the respective CCD beam profiles. Black dashed line is for an ideal gaussian mode for comparison. Solid dots indicate the grid size used for the respective Stark shift profiles, and the estimated power capture.

We use a bootstrapping method to determine the statistical uncertainty for $C$ in each case. The cubic spline to the data is taken as the mode function from which new datasets are generated by Monte Carlo with the addition of normally distributed variations in either (i) the stark shift measurements by the projection noise, (ii) an overall offset of the grid coordinates by up to 10 $\mu$m, (iii) or the coordinates of each point to account for positioning uncertainty. For profiles with a camera tracking the stage movement, the uncertainty of each coordinate position is taken to be $\pm 150$ nm. For the 1560, it is taken to be $\pm 1 \mu$m. In addition, linear drift of the coordinates at a rate of 9 $\mu$m/day in a randomly selected direction is also considered based on the findings in Sec. III. Table II gives the total uncertainty assessed for each in profile in Fig. IV with contributions from each of the sources considered.

TABLE II. Breakdown of uncertainty contributions to $C$ for each beam profile shown in Fig. IV.

| effect | 598 % | 646 % | 804 % | 848 % | 987 % | 1560 % |
|---|---|---|---|---|---|---|
| power capture | 0.2 | 0.1 | 1.1 | 1.6 | 1.5 | 1.1 |
| Stark measurements, projection noise | 0.5 | 0.28 | 0.5 | 0.9 | 0.4 | 0.24 |
| offset of grid coordinates | 0.3 | 0.08 | 0.3 | 0.32 | 0.35 | 0.09 |
| positioning statistical uncertainty | 0.06 | 0.15 | 0.11 | 0.09 | 0.11 | 0.35 |
| linear drift, thermal | 0.7 | 1.5 | 2.32 | 0.77 | 0.78 | 0.8 |
| total uncertainty | 0.95 | 1.6 | 2.7 | 2.0 | 1.8 | 1.4 |

For both 804 and 646 nm three profiles were taken, one of which from each set is show in Fig. IV. For 804, the three profiles yield $C$ values of 296.7(3.2), 293.2(5.1), and 294.1(8.0) mm$^{-2}$. The mean value found by $\chi^2$ minimization is 293.9 (2.6) mm$^{-2}$ with $\chi^2 = 1.33$ and $p$-value of 0.52. For 646, the three profiles yield 399.4(3.8), 401.4(5.0), and 397.2(6.2) mm$^{-2}$ with mean value of 399.5(2.7) mm$^{-2}$ for which $\chi^2 = 0.3$ with corresponding $p$-value of 0.87. Thus the profile measurements are repeatable to the within the estimated uncertainties. For other wavelengths a single profile was taken with the resulting normalizations reported in the main article.

## V. LASER SOURCES FOR STARK BEAMS

Here we give brief description of the laser sources and their frequency stability. Manufacturer model numbers are for reference only and not intended as a recommendation of any particular product.

1. **598 nm laser** : The laser source is a homebuilt external cavity diode laser (ECDL) using a gain chip (SAF1145-90-HTS) at 1196 nm. This is frequency doubled by a PPKTP waveguide doubler (ADVR Inc.). The laser is frequency stabilized to a Fabry-Pèrot cavity with a frequency offset. Every 2 seconds the frequency offset is adjusted to compensate the cavity drift and maintain the laser at fixed frequency with respect to an optical frequency comb. All 598 nm resonance frequencies were directly measured and found to be in agreement with the values reported in [1], where they had been inferred from closure of the 350 nm and 848 nm transitions.

2. **646 nm laser (polarizability method)** : A homebuilt ECDL using an Eagleyard (EYP-RQE-0650) gain chip. The laser is frequency stabilized to an aluminum spacer Fabry-Pèrot cavity which has approximately ±200 MHz stability over the day. The wavelength is measured by a WS8-10 wavemeter (± 10 MHz). All wavelength readings over the duration of the waist measurements were within the range $\Delta/2\pi = -241.7$ to $-242.3$ GHz. A diffraction grating is used to separate the amplified spontaneous spontaneous emission (ASE) from the laser carrier. With the grating filter the measured Stark shift increased by 2.9(6)% indicating the integrated ASE is only ∼16 dB less than the carrier. This laser is operated near to the edge of its gain profile and has particularly high levels of ASE compared the other diode laser sources.

3. **646 nm laser (scattering/stark ratio method)** : A homebuild ECDL using a HL638DG diode. Frequency stabilized to a reference transfer cavity in vacuum which is itself stabilized to the 848 nm clock laser. This 646 nm laser is used for laser cooling and detection of Lu$^+$ and has a frequency detuning $-199.46(10)$ MHz based on repeated measurements of the $|\,^3D_1, 7, 0\rangle \leftrightarrow |\,^3P_0, 7, \pm 1\rangle$ resonance frequencies. A frequency stability of ±10 kHz is observed when monitored against an optical frequency comb. Using the setup shown in Fig. VII, the light delivered to the experiment is offset to a detuning on the order of -1 GHz from the $|\,^3D_1, 7, 0\rangle \leftrightarrow |\,^3P_0, 7, 0\rangle$ resonance.

4. **804 nm laser**: Derived from the $^1S_0 \leftrightarrow\,^3D_2$ clock laser which is a homebuilt ECDL using an Axcel Photonics M9-808-150 diode. Frequency stabilized to a 10 cm ultra-low expansion (ULE) optical reference cavity with 30,000 finesse. Wavelength for the polarizability measurements was $804.125\,22\,(3)$ nm as measured by the WS8-10 wavemeter.

5. **848 nm laser:** Derived from the $^1S_0 \leftrightarrow\,^3D_1$ clock laser which is a homebuilt ECDL using a Thorlabs L852P150 diode. Frequency stabilized to a 10 cm ultra-low expansion (ULE) optical reference cavity with 400,000 finesse. Frequency shifted by an AOM to be several hundred MHz detuned from any clock transitions. Wavelength is $847.736\,132$ nm

6. **987 nm laser:** Toptica DL-Pro ECDL. Verified to be single mode on a Fabry-Pèrot cavity but not locked to a specific reference. The wavelength was monitored on a wavemeter (WS8-10) and did not drift by more

than ±700 MHz over the duration of the measurements. The wavelength was 987.086 (2) nm at the time the maximum Stark shift was measured for the polarizability assessment.

7. **1560 nm laser:** Homebuilt ECDL with Thorlabs SAF1126H gain chip which is amplified by a Nufern 5W Erbium-doped fiber amplifier. Laser frequency is stabilized to a high finesse reference cavity and the wavelength measured on a wavemeter (EXFO WA-1000, ± 300 MHz). This wavemeter was later found to be out of calibration by 1 GHz at 1762 nm compared to an atomic reference. The measured wavelength is 1560.80 (1)nm.

## VI. MAGNETIC FIELD ALIGNMENT

The following procedures are used to align the magnetic field to the $\phi = 0$, 54.7°, or 90° configuration.

**Alignment to $\phi = 0$:** Only for the 598 nm was the magnetic field aligned to $\phi = 0$ ($\pi$-coupling). This was achieved by tuning the laser to the $|\,^3D_1, 7, 0\,\rangle \leftrightarrow |\,^3P_1, 7, 0\,\rangle$ resonance and trimming the magnetic field to minimize the scattering rate out of $|\,^3D_1, 7, 0\,\rangle$. Compared to $\perp$-polarization, a suppression of the scattering rate by a factor of $10^4$ was observed at optimal alignment.

**Alignment to $\phi = 54.7°$:** To align the magnetic field to $\phi = 54.7°$, the magnetic field is rotated in the $yz$-plane until the tensor shift measured by the microwave clock transition is nulled. Note that even if there is small elevation angle to the magnetic field (tilt out of the $yz$-plane), zero tensor shift ensures the total angle between the magnetic field and linear laser polarization is correctly set to $\phi = \cos^{-1}(\sqrt{\frac{1}{3}}) \approx 54.7°$. In Sec. X, any residual angular misalignment is bound by the tensor shift measured on the microwave transition in each case.

**Alignment to $\phi = 90°$:** Here the magnetic field is rotated in the $yz$-plane to locate the extremum of the light shift. Fig. 7 shows data for alignment of the 646-nm laser for the polarizability measurement. The current, $I_z$ for the bias coil in the $\hat{z}$ direction is varied and the stark shift (red points) measured. Here, the stark shift is given by

$$\delta f = C - \frac{1}{2}C(3\cos^2\phi(I_z) - 1),$$

with

$$\phi(I_z) = \left(\frac{\pi}{2} - \alpha\right) - \tan^{-1}\left(\frac{\sin(\theta) + KI_z}{\cos(\theta)}\right),$$

where $C$, $\theta$, $K$, and $\alpha$ are fit parameters. From the fit shown, the optimal angle is located with an estimated uncertainty of 2 mrad.

The method assumes the magnetic field is in the $yz$-plane. The elevation angle of the magnetic field is set adjusting the $\hat{x}$ component of the magnetic field to optimize $\pi$-coupling for the linearly polarized optical pumping beam propagating along $\hat{x}$. An error in this angle could arise in as much as the stark laser and optical pumping laser are not orthogonal. Due to the geometry of the vacuum chamber, it is unlikely these beams are more than two degrees from perpendicular. An angular misalignment of $\gtrsim 5°$ would be required for 1% error in the reported polarizabilities.

FIG. 7. Alignment of the magnetic field perpendicular to the laser polarization

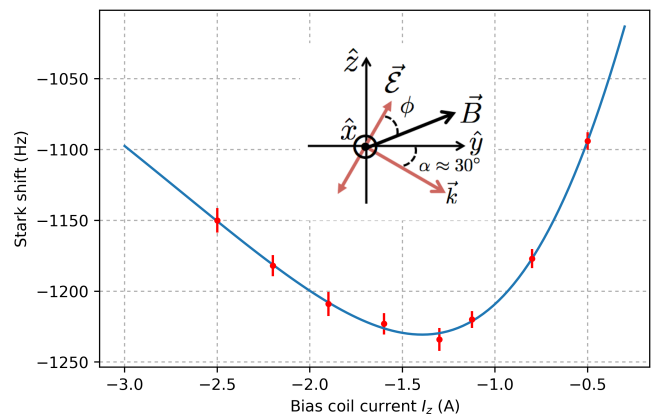

## VII. 646 SCATTERING TO STARK SHIFT RATIO DATASET

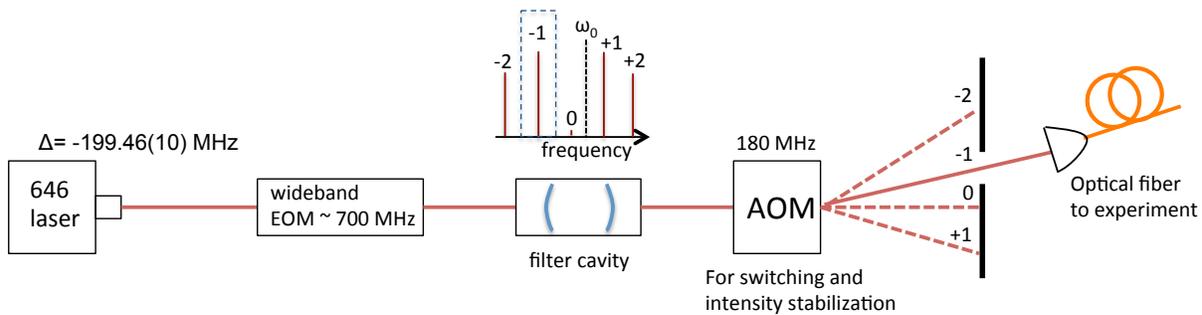

FIG. 8. Setup of 646 nm laser for scattering to stark shift ratio measurement. A filter cavity transmits only the -1 order sideband of the EOM.

The setup of the 646 nm laser is shown schematically in Fig. VII. A wideband electro-optic phase modulator (EOM) is operated at modulation depth sufficient to eliminate the carrier. A ~2000 finesse cavity of length 2 mm transmits only the -1 order sideband generated by the EOM. Finally, the -1 order of an AOM used for switching and active stabilization of the laser intensity is sent to the experiment by optical fiber. In an earlier setup which did not have the EOM and filter cavity, it was found the scattering rate was increased by as much as 20% with the AOM operated near 200 MHz (Fig. VII). We attribute this to spurious resonant light, which we speculate is from a -2 order sideband from residual amplitude modulation by the AOM at the rf frequency. A spurious sideband at the level $\sim -50$ dBc is sufficient to explain the outliers in Fig. VII. The setup in Fig. VII avoids this problem by shifting the laser to a larger detuning ($\sim 1$ GHz) to avoid unwanted resonant components generated by the AOM. The filter cavity acts not only to suppress the unwanted EOM sidebands, but also ASE or any spurious components which are near resonant. In this configuration, preliminary measurements of the matrix element were taken for several EOM frequencies spanning 600 to 700 MHz, spanning a detuning range $\Delta_0/2\pi = -979.46(10)$ to -1079.46(10) and no correlation with frequency was observed. Much longer datasets to test the stability and repeatability of the measurement were taken at two detunings, -989.46(10) and -1119.46(10) MHz, to arrive at the results reported in the main article. In this section, we present the raw data and statistical analysis from one of these datasets at $\Delta_0/2\pi = -989.46(10)$.

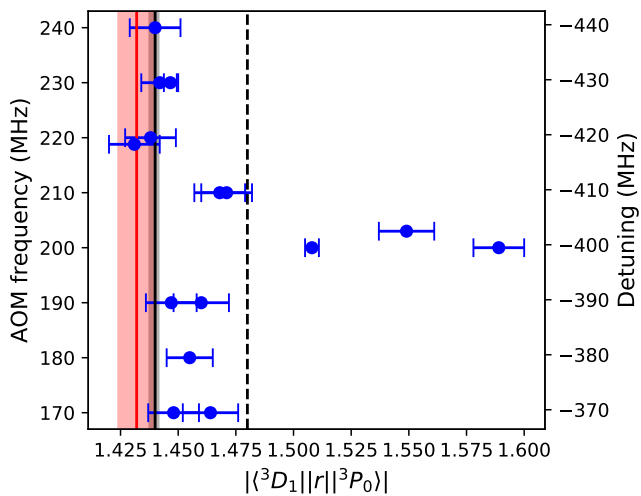

FIG. 9. Without the EOM and filter cavity in Fig. VII, results are contaminated by excess scattering from a spurious resonant sideband when the AOM is operated near 200 MHz. The vertical lines are the results reported in the main article Fig. 3b for reference.

As described in the main article, the dataset consists of three interleaved experiment: (i) detection immediately after state preparation to measure $P_S$ (ii) pulse the 646 laser for $\tau = 25$ ms, shelve $|{}^3D_1, 7, 0\rangle$ population to $|{}^1S_0, 7, +1\rangle$,

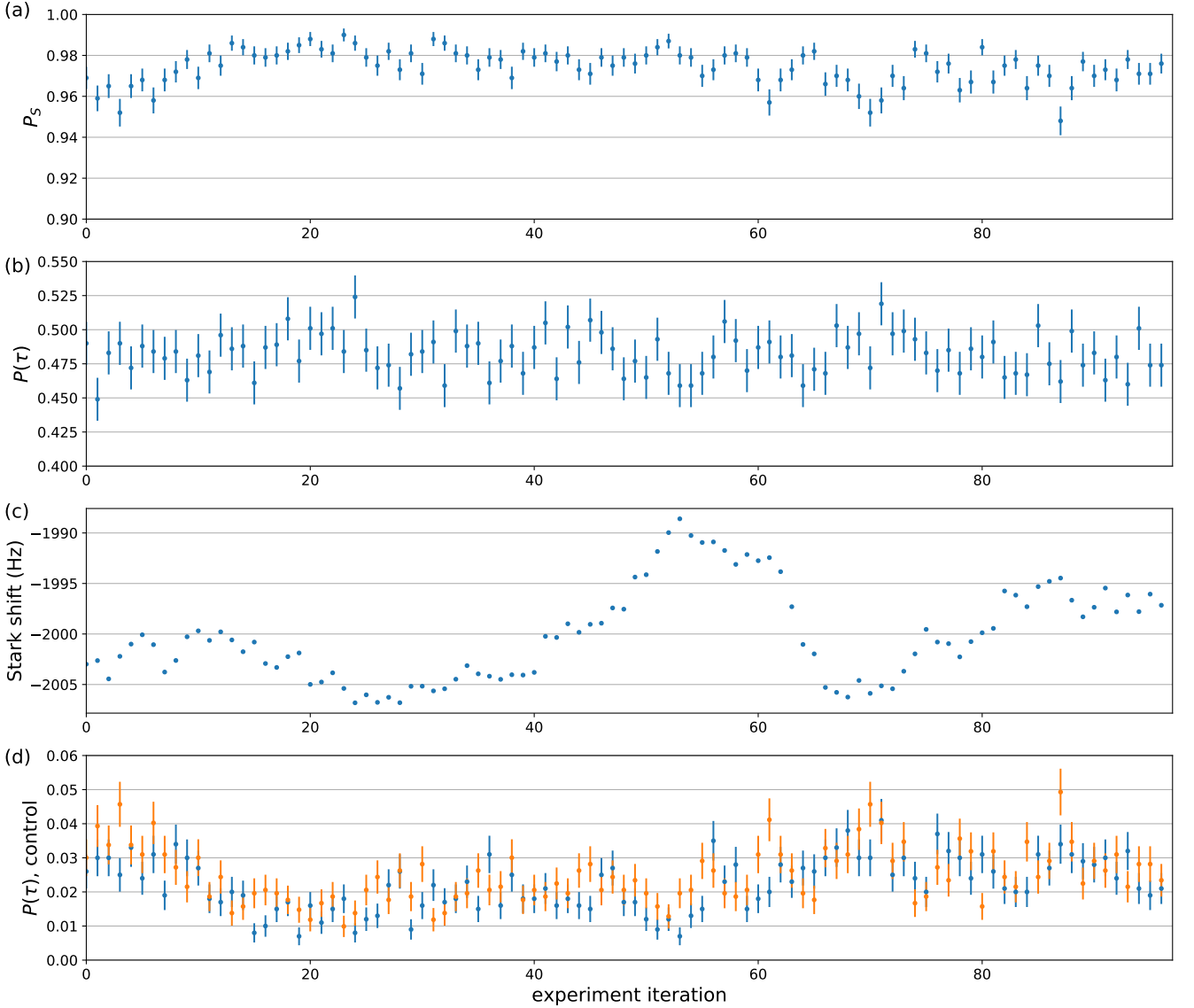

FIG. 10. (a-d) results of interleaved experiments (i) to (iv). The orange points in (d) are $P_S(1 - P_S)$ computed from (a).

and measure the remaining bright population $p(\tau)$, and (iii) measure the stark shift induced by the 646 laser using an alternating clock servo with $\pi$ time of 6 ms. In addition there was a fourth control experiment which was not mentioned in the main text: (iv) the same sequence as (ii) but without pulsing the 646 laser during the 25 ms between preparation and shelving. The control experiment should be equal to $P_S(1 - P_S)$ as long as there is no significant scattering out of the $|^3D_1, 7, 0\rangle$ state during the 25 ms from unwanted sources, such as leakage light from the near resonant 646-nm cooling and detection lasers. Fig. 10 shows the results of experiments (i) to (iv) where each point represents 1000 experiment iterations. The acquisition time for the entire dataset was 15 hours. All error bars shown are projection noise limited uncertainties. The orange points in Fig. 10d are $P_S(1 - P_S)$ computed for the data in Fig. 10a which are clearly correlated with the control experiment results (blue points). This confirms the drift in the control experiment is entirely due to slow variation in the clock shelving probability $P_S$ as expected.

Using the model described in the main text, the matrix element is determined from $P_S$, $p(\tau)$, and the Stark shift $\delta_0$ for each 1000 iteration block (Fig. 11a). The mean value (solid line) of $|\langle^3D_1||r||^3P_0\rangle| = 1.441\,(33)$ has reduced $\chi^2 = 0.87$ with $p$-value of 0.81. The allan deviation (ADEV, Fig. 11b) also indicates the result is averaging down in accordance with projection noise limited statistics (black line). The total deviation (TOTDEV) is additionally given as a better estimator of the stability at long averaging times [4]. This result is one of the square points reported in Fig. 3b of the main article.

FIG. 11. (a) Matrix element estimated from each block of 1000 experiments. Blue solid line is the mean with reduced $\chi^2 = 0.87$. (b) Allan (total) deviation showing the stability of the measurement.

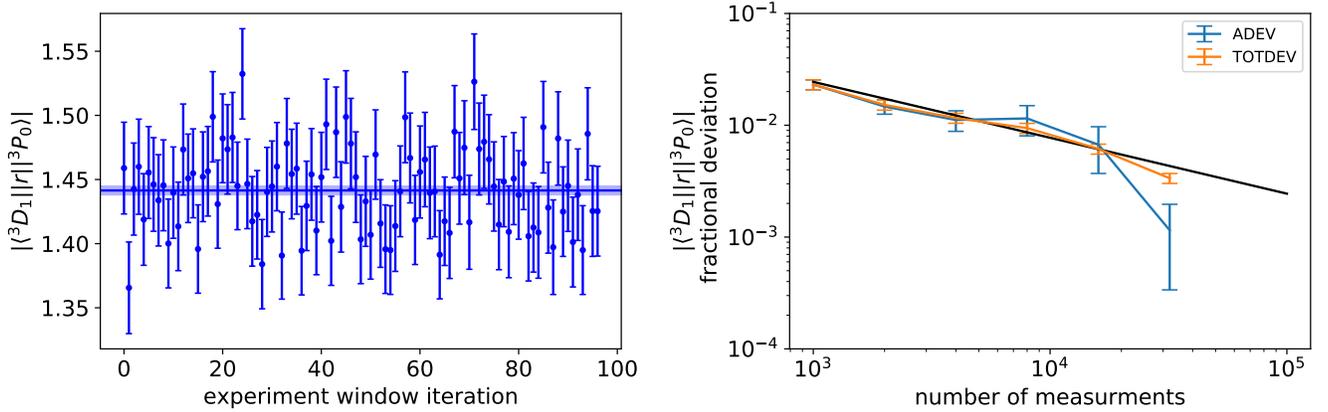

## VIII. $^3P_1 \to {}^3D_1$ BRANCHING RATIO

FIG. 12. Branching ratios from $^3P_1$ $F=7$.

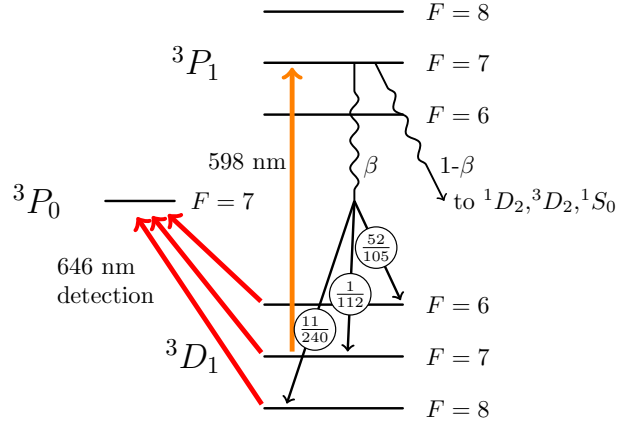

To measure the $^3P_1 \to {}^3D_1$ branching ratio, $\beta$, we use the 598 nm laser to resonantly pump population out of the $^3D_1$ $F=7$ manifold and measure the resulting population in the $^3D_1$ $F=6$ and $F=8$ manifolds. The 598 nm laser is linearly polarized perpendicular to the external magnetic field. The laser frequency is tuned to the $|{}^3D_1, 7\rangle \leftrightarrow |{}^3P_1, 7\rangle$ resonance.

We use a field programmable gate array (FPGA) to implement a real-time Bayesian detection algorithm to determine if the atom is in bright ($^3D_1$) state [2, 3]. Detection errors are not symmetric: if the ion starts in a dark state it will never be detected bright, but if the ion starts in the bright state it will sometimes be detected dark due to decay out of the detection channel before the sufficient photons have been collected to determine the ion is in the bright state.

At the start of every experiment, a cycle of repumping and detection is repeated until the ion is successfully detected bright. If this is immediately followed by another detection event, the ion is detected bright with probability $P_b = 0.9990(1)$. This is the measured detection efficiency of the bright state, limited by the aforementioned state detection errors.

To measure the branching ratio, after a successful bright detection a 20 $\mu$s pulse of 646 nm light resonant with the $|{}^3D_1, 6\rangle$ and $|{}^3D_1, 8\rangle$ to $|{}^3P_0, 7\rangle$ transitions pumps all population into the $|{}^3D_1, 7\rangle$ hyperfine manifold. This is

followed by a 598 nm laser pulse of 400 $\mu$s which pumps all population out of the $|\,^3D_1, 7\,\rangle$ manifold. Fig. 12 shows the decay paths from $^3P_1$. The final population pumped into the $|\,^3D_1, 6\,\rangle$ and $|\,^3D_1, 8\,\rangle$ manifolds, $P_f$, is related to $\beta$ by

$$P_f = \frac{\frac{52}{105}\beta + \frac{119}{240}\beta}{1 - \frac{1}{112}\beta}. \tag{1}$$

After the 598 nm pulse, we measure a bright population of $0.1846\,(13)$ from $8.8 \times 10^5$ experiments. This is scaled by the bright detection efficiency $P_b$ to find $P_f = 0.1848(13)$. The branching ratio from Eq. (1) is $\beta = 0.1862(13)$. A value of $0.1862\,(17)$ was previously measured in $^{175}$Lu$^+$ [2] and the theory value is 0.186 [2].

## IX. 598 NM SCATTERING RATIO TO STARK SHIFT DATASET

In this section one of the four datasets for the 598 nm scattering to Stark shift ratio measurement is presented. This dataset is taken at detuning of $\Delta_1/2\pi = 995.7(1)$ MHz from the $|\,^3D_1, 7, 0\,\rangle \leftrightarrow |\,^3P_1, 8, 0\,\rangle$ transition. The conditional state preparation sequence describe in the main article is used to prepare population in $|\,^3D_1, 7, 0\,\rangle$ for the scattering rate measurement. Three experiments are interleaved: (i) detection immediately after state preparation to measure $P_0$ (ii) pulse the 598 laser for $\tau = 30$ ms and measure remaining bright population $p(\tau)$, and (iii) measure the Stark shifts induced by the 598 laser using an alternating clock servo with $\pi$ time of 5 ms. From $5 \times 10^4$ experiments, we measure $P_0 = 0.9880(6)$. Fig. 13 shows the data for $5 \times 10^4$ measurements of $p(\tau)$ and for $1.6 \times 10^5$ clock interrogations, alternately with and without the 598 laser, to measure $\delta_1$.

FIG. 13. Measurement data from 598 nm scattering rate and stark shift.

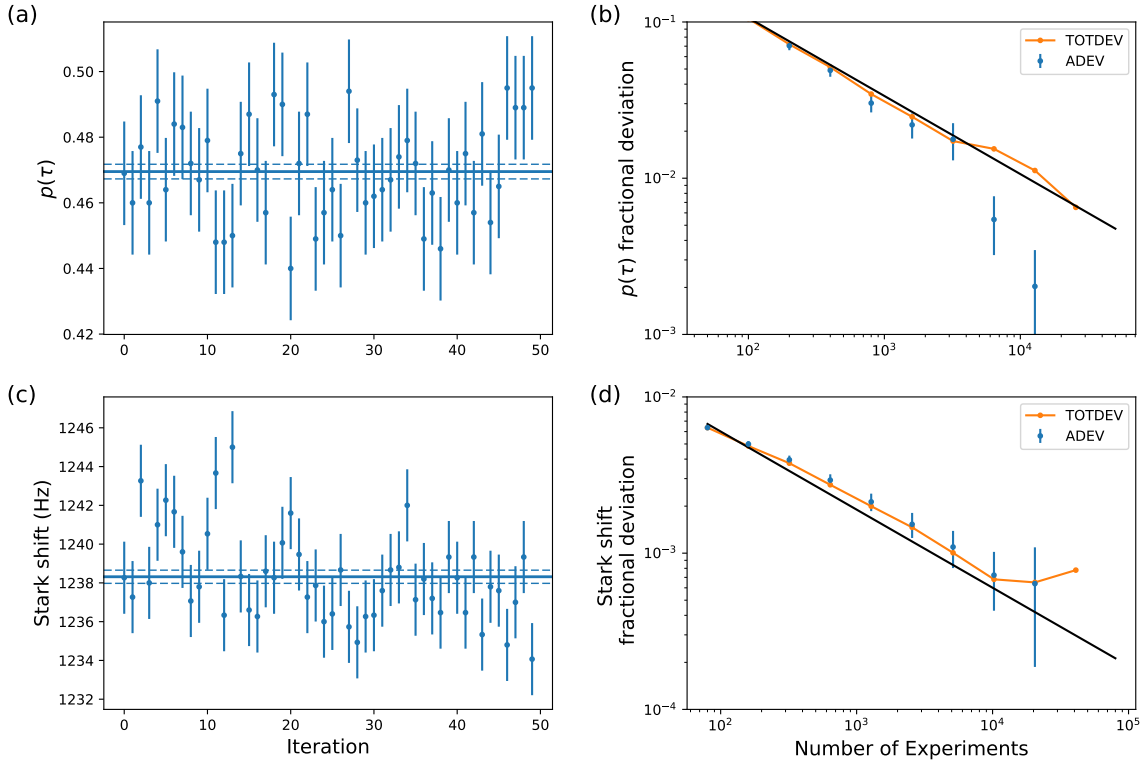

In Fig. 13a each point represents the average of 1000 experiments measuring $p(\tau)$. The mean value is $p(\tau) = 0.470(2)$ with a reduced $\chi^2 = 0.89$ and $p$-value of 0.70. Fig. 13b shows the fractional allan of $p(\tau)$ where the black line is the projection noise limit. Both the $\chi^2$ and allan deviation are consistent with projection noise limited statistical uncertainty.

In Fig. 13c each point represents 1600 clock interrogations to measure the Stark shift with error bars assuming projection noise limited measurements. The mean value (solid line) has reduced $\chi^2 = 1.6$ with a $p$-value of 0.004, indicating an unaccounted for systematic uncertainty. The allan (total) deviation in Fig. 13d also indicate an instability at the fractional level of $4 \times 10^{-4}$ of the stark shift. Mostly likely this is the limit of the optical power stability for the 598 laser due to either the intensity servo or the beam pointing stability. Adding a $4 \times 10^{-4}$ fractionally systematic uncertainty (0.5 Hz) to each point in Fig. 13c brings the $\chi^2$ to 1. We take this as systematic uncertainty for mean of the Stark shift with the result $\delta_1 = 1238.3(5)$ Hz. In any case, the uncertainty in the final result is not limited by $\delta_1$.

Scattering via the $|\,^3P_1, 8, 0\,\rangle$ state instead of $|\,^3P_1, 6, 0\,\rangle$ state requires straight forward modification to the model presented in the main article to account for the different Clebsch-Gordon coefficients. For the scattering rate

$$R_1 = \Gamma_1 \frac{7}{90} \frac{\Omega_1^2}{4\Delta_1^2}, \qquad (2)$$

the bright population is given by

$$p(\tau) = P_0 \left[ \frac{9\beta}{16 - 7\beta} + \frac{16 - 16\beta}{16 - 7\beta} e^{-R_1(1-\frac{7}{16}\beta)\tau} \right], \qquad (3)$$

and the stark shift is given by

$$\delta_1 = \frac{\Omega_1^2}{4} \left( \frac{4}{45} \frac{1}{\Delta_1 + \omega_{68}} + \frac{7}{90} \frac{1}{\Delta_1} \right). \qquad (4)$$

Solving the equations to eliminate $\Omega_1$, we find $\Gamma_1 = 2\pi \times 4.255(28)$ MHz and

$$|\langle\,^3D_1||r||^3P_1\rangle| = 1.257(7). \qquad (5)$$

This result is represented by the blue circle in Fig 4b of the main article.

## X. STARK SHIFT DATA FOR POLARIZABILITY RESULTS

The differential ac-Stark shifts induced on the $|\,^1S_0, 7, 0\,\rangle \leftrightarrow |\,^3D_1, 7, 0\,\rangle$ optical clock transition, $\delta f$, and $|\,^3D_1, 6, 0\,\rangle \leftrightarrow |\,^3D_1, 7, 0\,\rangle$ microwave transition, $\delta f_\mu$, due to linearly polarized laser light of frequency $\omega$ are given by [5]

$$\delta f = -\frac{1}{2h} \langle E^2 \rangle \left( \Delta\alpha_0(\omega) + \frac{1}{2}\alpha_2(\omega)(3\cos^2\phi - 1) \right) \qquad (6)$$

$$\delta f_\mu = -\frac{1}{2h} \langle E^2 \rangle \left( \frac{7}{10}\alpha_2(\omega)(3\cos^2\phi - 1) \right) \qquad (7)$$

where $\phi$ is the angle between the laser polarization and quantization axis and $\langle E^2 \rangle$ is mean squared electric field averaged over one optical cycle. This is related to the laser intensity by $\langle E^2 \rangle = I_0/(c\epsilon_0)$. The peak laser intensity is $I_0 = CP_0$ where $P_0$ is the power at the ion and $C$ is the normalization coefficient determined from the beam profile. Polarizabilities are reported in atomic units which can be converted to SI units via $\alpha/h\,[\mathrm{Hz\,m^2\,V^{-2}}] = 2.48832 \times 10^{-8} \alpha$ (a.u.).

To find $\Delta\alpha_0(\omega)$, the magnetic field is rotated to $\phi_m = \cos^{-1}(\sqrt{\frac{1}{3}}) \approx 54.7°$ where the tensor contribution to $\delta f$ is nulled. The optimal angle is found by measuring $\delta f_\mu$. Table III gives measured values of $\delta f$ and $\delta f_\mu$ at the optimized angle. The residual Stark shifts measured on the microwave transition imply $\phi$ has been set to within 1 mrad of $\phi_m$ for every wavelength. Uncertainties on all Stark measurements are statistical from projection noise. To find $\alpha_2(\omega)$, the field is rotated to find the extremal Stark shift near $\phi = \pi/2$ by the procedure described in Sec. VI. The measured shifts on the microwave transition at this position are given in Table III.

TABLE III. Measured ac-Stark shifts used to determine the polarizabilties.

| wavelength (nm) | $P_0$ (mW) | C (mm$^{-2}$) | $\delta f(\phi_m)$ (Hz) | $\delta f_\mu(\phi_m)$ (Hz) | $\delta f_\mu(\phi = \pi/2)$ (Hz) |
|---|---|---|---|---|---|
| 804 | 12.49(25) | 293.9(2.6) | -316.0(0.3) | -0.01(0.26) | 168.3(0.1) |
| 848 | 16.19(17) | 268.3(5.3) | -286.1(0.2) | 0.13(0.11) | 165.2(0.1) |
| 987 | 29.18(26) | 271.4(4.8) | -280.8(0.4) | -0.07(0.42) | 209.2(0.1) |
| 1560 | 441 (10) | 84.2(1.2) | -386.0(0.3) | -0.45(0.37) | 698.4(0.1) |



\end{document}